 \documentclass[preprint2,longabstract]{aastex}

\slugcomment{To be submitted to ApJ on 7 July 2006}

\shorttitle{The pc-scale structure and evolution of BL Lac PKS~0003$-$066}
\shortauthors{Horiuchi et al.}

\begin{document}


\title{Evolution of the parsec-scale radio structure in the BL Lac object \\
PKS~0003$-$066}


\author{S. Horiuchi, 
S.J. Tingay, C.J. West
}
\affil{
Centre for Astrophysics and Supercomputing,
Swinburne University of Technology,\\
Mail No. 39, P.O. Box 218, Hawthorn, Victoria 3122, Australia
}
\email{shoriuchi@swin.edu.au
}

\author{A.K. Tzioumis, C.J. Phillips, J.E.J. Lovell, J.E. Reynolds 
}
\affil{
Australia Telescope National Facility, CSIRO,
P.O. Box 76, Epping, NSW 1710, Australia
}

\author{S.P. Ellingsen, P.B. Reid, G. Cim\`{o} 
}
\affil{
School of Mathematics and Physics, University of Tasmania,\\
Private Bag 21, Hobart, Tasmania 7001, Australia
}

\author{J.F.H. Quick 
}
\affil{
Hartebeesthoek Radio Astronomy Observatory,\\
P.O. Box 443, Krugersdorp 1740, South Africa}

\author{Y. Koyama 
}
\affil{Kashima Space Research Center,
National Institute of Information and Communications Technology,
893-1 Hirai, Kashima, Ibaraki 314-8501, Japan}

\author{F.H. Briggs 
}
\affil{
The Australian National University, Mount Stromlo Observatory, Australia, \\
Australia Telescope National Facility, CSIRO,
P.O. Box 76, Epping, NSW 1710, Australia}


\author{W.F. Brisken 
}
\affil{National Radio Astronomy Observatory, P.O. Box O, Socorro, NM 87801}




\begin{abstract}

We present the results of global VLBI observations of
the BL Lac object PKS~0003$-$066 at 2.3 and 8.6 GHz, over a 10 year 
period using data from new and archived VLBA, geodetic VLBI, and global 
VLBI (including the use of new disk-based recording and software correlation systems) 
observations. Inter-continental baselines resolve this source into a compact 
core and three jet components that move away from the core with proper 
motions ranging between $\sim$0.1 mas/yr 
for the outer jet component and $\sim$1.4 mas/yr for the fastest inner 
jet component, i.e. apparent speeds of $\sim$2.0 $c$ to $\sim$21.6 $c$ respectively.  
This result is in contrast to previous studies of this source which did not 
reveal significant jet motions from limited data sets.  
We find that the fast inner jet components catch up to the more slowly moving 
outer jet components and interact strongly, causing local brightening of 
the jet in the interaction.  We document the body of VLBI data for this source, 
as well as other supporting radio observations, as a basis for future work that will 
investigate the underlying physics of the pc-scale jet in PKS 0003$-$066.
    
\end{abstract}





\section{Introduction}

BL Lac objects are extragalactic sources characterised by
weak or undetectable optical line emission and strong and variable
flux and polarization in a wide range of wavebands from optical through
radio (for review, e.g. Kollgaart 1994). 
In the radio band, very long baseline interferometry 
(VLBI) images show that the pc-scale
structures of BL Lac objects are dominated by compact, flat-spectrum cores
from which (apparently) one-sided jets emerge, often showing knots of
enhanced brightness with apparent superluminal speeds 
(e.g Gabuzda, Pushkarev \& Cawthorne 2000).
These properties are explained by Doppler boosted synchrotron radiation
from a relativistic jet closely aligned to
our line of sight (Blandford \& Rees 1974).
In the framework of the unified model
of radio-loud active galactic nuclei (AGNs)
(see e.g., Urry \& Padovani 1995), the observed
differences in the properties of quasars, radio galaxies,
and BL Lac objects are due to the orientation of the
relativistic beam and an obscuring torus with respect to
our line of sight.

The BL Lac object PKS~0003$-$066 (NRAO~005, J0006$-$0623) at a redshift of
$z$ = 0.347 (Stickel, Fried \& K\"{u}hr 1989) is a bright compact radio source
and has been observed with VLBI 
for many projects (e.g. Fey \& Charlot 1997).
The core is unresolved
at arcsecond resolution (Ulvestad et al. 1981, Perley 1982).
Gabuzda, Pushkarev \& Cawthorne (2000) observed PKS~0003$-$066 with a global 
VLBI array and showed that 
the pc-scale structure is resolved, with the major components strongly polarised, 
but couldn't identify any superluminal motions from comparison to previous VLBI 
observations. Kellermann et al. (2004) analysed the results
of a 15 GHz (2 cm) multiepoch Very Long Baseline Array (VLBA) program
begun in 1994, to study the jets in 110 quasars and active galaxies. 
PKS~0003$-$066 is one of 18 BL Lac objects in that sample,
Kellermann et al. (2004) identifing three jet components with no
significant proper motions.  
Total flux density monitoring of this source with the ATCA at 1.4, 2.5, 4.8 
and 8.6 GHz shows rapid variability over a 2 - 3 year period,
with a higher degree of variability at higher frequencies (Tingay et al. 2003),
suggesting variability of the pc-scale structures.

To investigate this possibility,
we have made new global VLBI observations of PKS~0003$-$066 at 2.3 GHz.
We have also analysed 25 other VLBI observations at 2.3 and 8.6 GHz
from a 10 year series of observations to study the flux density and structural 
variability of the source in more detail than has previously been possible.
In this paper we describe these results and a comparison to the results 
of previous multi-frequency and multi-epoch VLBI experiments.
In contrast to the results in Kellermann et al. (2004),
we find evidence for fast jet components ejected from the core which 
interact strongly with slower jet
components further from the core and cause local jet brightening.

In section 2.1, we describe new VLBI observations at 2.3 GHz 
using a global VLBI array, 
new disk-based recording techniques, and software correlation techniques.
 In section 2.2 we summarize the results obtained from 
archival VLBI data. In section 3 we discuss the pc-scale structure and 
evolution of this source.  

Throughout this paper we use the following cosmological parameters
as used by Kellermann et al. (2004): 
$H_0$ = 70 kms$^{-1}$Mpc$^{-1}$, $\Omega_{m} = 0.3$, $\Omega_{\Lambda} = 0.7$.  
The corresponding luminosity distance is 1.939 Gpc 
at the redshift of PKS~0003$-$066, $z = 0.347$, and 
the linear scale is therefore 4.91 pc/mas.
A proper motion of 1 mas/yr corresponds to an apparent speed 
of $16c$.

\section{Observations and Results}
\subsection{Global VLBI observations at 2.3 GHz}

\subsubsection{Global disk-based VLBI observations}

PKS~0003$-$066 was observed with a global array of antennas on 2004 April 18, 
using 5 Australian telescopes, one South African telescope (Hartebeeestoek Radio Astronomy Observatory), 
and one Japanese telescope (Kashima),
as summarised in Table 1, for a period of 20 minutes.  This experiment
was designed as a test observation for the software correlator described
in section 2.1.2.
The observational setup utilised a number of different disk-based
recording systems (as briefly summarised in Table 1) to record 
the Nyquist-sampled and digitised telescope output as a function of time.
At the antennas that utilised the LBA Data Recorder (LBADR)
\footnote{The LBADR recorders use the VLBI Standard Interface I/O Board
(VSIB) and the VLBI Standard Interface Converter(VSIC) cards 
developed by the Mets\"{a}hovi Radio Observatory 
(Ritakari \& Mujunen 2002), along with Apple XRaid units for data storage, 
except Ceduna which used internally raided drives for this experiment.  
For further details, see Phillips et al. (2006, in preparation)}
recording system (West 2004), an aggregate data rate of 256 Mbps was recorded, 
corresponding to four 16 MHz bands, each
sampled at 32 MHz and 2-bit digitised.  The four 16 MHz bands were
arranged with two bands at right circular polarisation (RCP) (2274 - 2290 MHz) 
and two bands at left circular polarisation
(LCP) (2274 - 2290 MHz). Duplicate bands were recorded at RCP and LCP 
for redundancy.
At the antennas that utilised the Mark5 recording system, an aggregate
data rate of 128 Mbps was recorded, in matching 16 MHz bands.
At the Kashima antenna, which utilized the Japanese K5 recording
system (Koyama et el. 2003), an aggregate data rate of 1 Gbps was recorded.
The 2274-2306 MHz frequency range of the RCP signal was converted 
to a 0-32MHz baseband signal, sampled at the rate of 512 MHz
and then digitized with 2-bit precision. The 1 Gbps data was digitally processed to
obtain the time series data corresponding to 
a single 16 MHz band sampled at 32 MHz using 2-bit digitization.
This digitally filtered 16 MHz band ranged from  2274 - 2290 MHz, 
corresponding to one of the 16 MHz bands recorded using the LBADR 
and Mark5 systems at the other antennas.  
An account of the software used to perform the digital
filtering can be found in Takeuchi (2004).

This experiment illustrates the power and flexibility of disk-based
recording for VLBI.  
Traditional tape-based VLBI requires considerable effort to translate 
different recorder systems prior to correlation. Our software correlator 
is capable of reading different disk-based recorder formats without 
conversion, improving the compatibility of antennas with 
different recording systems.
The data recorded at each antenna were shipped to the Swinburne University
of Technology, for correlation using the software correlator described in
section 2.1.2.  Since the Kashima
antenna only produced a single 16 MHz band (RCP) using the software digital
filter, only this band was correlated with a view to full data reduction
and imaging.

\subsubsection{Description of correlation in software}

A software correlator for VLBI has been developed using the Beowulf
cluster at the Swinburne University of Technology Centre for Astrophysics
and Supercomputing.  The Beowulf cluster consists of a large number
($\sim$300) of Pentium based PCs 
(for detail, see http://astronomy.swin.edu.au/supercomputing/).

The software correlator is an $XF$ style correlator
\footnote{This is a proto-type software correlator (West 2004). 
The correlator has more recently been superceded with a more capable 
$FX$ software correlator (Deller et al. 2006 in preparation).}.  
Data recorded to hard disk 
at each individual telescope are mounted 
on the supercomputer and the data for each antenna pair (baseline) is transfered 
to a processing node for correlation, in segments of up to 25 ms at a
time.  High level control of the data flow is provided under the standard
Message Passing Interface (MPI).  The data to be transfered is selected
from the data on disk according to the geometric delay model and antenna
clock models at the selected time in the observation; the geometric delay model 
is based on an imaginary telescope located at the centre of the Earth, 
allowing the two data streams to be aligned to +/- 1 sample accuracy (the fractional
sample error is retained for use in post-correlation corrections).  
The geometric delay model is generated using the CALC package 
(Himwich 1988, see also http://gemini.gsfc.nasa.gov/solve/solve.html). 

Once the selected data from two antennas (a single baseline) are
transfered to a processing node for correlation, the data are unpacked into
32-bit floating point numbers.  The fringe rotation function
derived from the predicted delay between the two antennas is then formed
and applied to one of the two data streams.  The two data streams are then
cross correlated to form the lag spectrum (expressed in terms of 
the correlator coefficient at each lag), and the lag spectrum Fourier
transformed to give the cross power spectrum.  In order to recover the
full signal to noise (e.g. D'Addario 1989), a second fringe rotation
function is generated, $\pi/2$ out of phase from the first fringe rotation
function.  The correlation is repeated using the second fringe rotation
function.  The two resulting cross power spectra are then coherently
averaged, after correction for the $\pi/2$ phase shift, to give the final 
cross power spectrum.  A fractional sample error correction is applied to 
the cross power spectrum, to correct for the phase slope across the band 
due to the limited accuracy in the initial alignment of the two data streams.
Finally, the measured system temperatures and gains from each antenna are 
applied to calibrate the visibility amplitudes into Janskys (see Table 1). 

The correlated data are reported back to a master node for collation and are 
exported to the FITS format, for further reduction in
standard processing packages.  This correlation scheme closely follows that
of the Australian Long Baseline Array  
(LBA) correlator (Wilson et al. 1992; Roberts et al. 1997).
An account of a prototype version of the software correlator can be found 
in West (2004). This new system for global VLBI is available to all users 
through the Australia Telescope National Facility (ATNF)
(for detail, see http://www.atnf.csiro.au/vlbi/).

\subsubsection{Post-correlation data reduction and imaging analysis}

Following correlation, the data were imported into the AIPS
package for fringe-fitting and phase calibration, 
then into the DIFMAP package (Shepherd 1997) for editing,
imaging, and evaluation of the source structure using modelfitting in the
($u,v$) plane.
The ($u,v$) coverage for the observation is thus shown in Fig. 1.  
The coverage is reasonably good for a 20 minute observation (essentially 
a single snapshot) due to the long east-west baselines from Australia to
South Africa, the long north-south baselines from Australia to Japan, and
the longest baseline from Japan to South Africa, which has large
components in both the east-west and north-south directions.
Fig. 2 shows the visibility amplitude as a function of distance in the
($u,v$) plane for the observation, projected onto the $u$ axis.  Structure in
the source is immediately apparent from the variations in the visibility
amplitude.

The data were model-fitted in the ($u,v$) plane using a three component model
(Table 2) and self-calibrated in amplitude and phase.  Amplitude
corrections of less than 5\% were derived from this procedure, showing
that the {\it a priori} calibration (Table 1) was of a reasonable quality.  
The image resulting from the modelfit is shown in Fig. 3.  
As shown in section 2.2
and discussed in section 3, the structure of PKS~0003$-$066 derived from
these observations agrees well with the structure found from other VLBI
array/correlator combinations such as VLBA+geodetic antennas  
at 2.3 and 8.6 GHz (Fey \& Charlot 1997, see also below), the VLBA at 5 GHz 
(Fomalont et al. 2000, Gabuzda, Pushkarev \& Cawthorne 2000), 
and the VLBA at 15 GHz (Kellermann et al. 1998).
The core component is marginally resolved on the Australia-Japan-South Africa 
baselines, which are the longest baselines yet used to observe this source, 
apart from the VSOP observation for this source where three ground telescopes were used 
for two hours in conjunction with the HALCA space telescope (Scott et al. 2004).  
The brightness temperature for the core component, $T_b$, is derived to be 
5.3$\times 10^{11}$K from Table 2, being consistent with the lower limits of 
$>4.8\times 10^{11}$K derived by Scott et al. (2004) from the VSOP observations 
(see also Horiuchi et al. 2004) and $>4.60\times 10^{11}$K derived by Kovalev et al. 
(2005) from a VLBA 15 GHz observation. 

\subsection{Analysis of VLBA/geodetic archive data}


Additional to the data described above, to study the structural variability of PKS~0003$-$066, we have used 
data from the Radio Reference Frame Image Database (RRFID) of the 
US Naval Observatory (USNO), at 2.3 and 8.6 GHz (Fey \& Charlot 1997).
Observations were made using an array consisting of 
the 10 antennas of the Very Long Baseline Array (VLBA) 
of the National Radio Astronomical Observatory (NRAO) 
along with geodetic antennas. In total 13 epochs of data
are available from 1995 October to 2002 January (see Table 3).
Each observation was undertaken with the
dual-frequency VLBA receivers.
The sources were observed using short duration ($\sim$ 3 minutes)
``snapshots" over a number of different hour angles.
We analysed and imaged using DIFMAP the original datasets for PKS~0003$-$066 
kindly provided to us by Alan Fey (private communication).  
As images from the RRFID are
already published 
\footnote{All images are available from the USNO-RRFID 
website (see http://rorf.usno.navy.mil/RRFID/)
and the NASA/IPAC Extragalactic Database (NED)}, 
we do not reproduce them in full here.  
Full results of our model-fitting
analysis of the data are presented, however, in Table 4
for observations listed in Table 3. 

We have also obtained two datasets from the VLBA online archive for 
PKS~0003$-$066. One is a part of a series of coordinated 
RDV (Research and Development VLBA)
astrometric/geodetic experiments, numbering in total 9 epochs from 2002 
July to 2004 July.  These experiments use the full 10-station VLBA plus 
up to 10 Mark 4 geodetic stations currently capable of recording VLBA modes. 
These experiments were coordinated by the geodetic VLBI programs 
of three agencies: USNO, NASA, and NRAO to monitor source structure 
and to determine a high accuracy terrestrial reference frame (Gordon 2000).

Another dataset consists of three epoch observations at 2.3 and 8.6 GHz 
with the VLBA only (obs. code BP 118) in May, June, and July 2004.
These archival data are exported to AIPS and DIFMAP for imaging 
using standard procedures.
Since these images from the VLBA archive data have not been previously presented, 
in Figs. 4, 5 and 6 we show images obtained from our analysis of the 9 epoch 
RDV data and the 3 epoch BP 118 VLBA data. 
Parameters of the images are given in Table 5.

We thus have a total of 25 epochs of data at 2.3  and 
8.6 GHz in addition to the global VLBI observation at 2.3 GHz
described in Section 2.1. 
Figures 7 and 8 show subsets of the RRFID and RDV images as montages at 
both 2.3 and 8.6 GHz over a 10 year period.
The image from year 2003.35 is from the RDV data, 
images at all other epochs are from the RRFID data.

From a model-fit analysis of the 8.6 GHz data we identified 3 components 
in the pc-scale jet (as did Kellermann et al. 2004) that are 
persistently recognisable until 1998,
in addition to the core. Then a new, 4th, component emerged from
the core in late 2002. 
Table 4 lists the source models we obtained at each epoch. 
Fig. 9 shows the angular distance of each component from the core as 
a function of time.  Fig. 10 shows the offsets of the jet 
components in right ascension and declination
from the core position, as a function of time.  
Finally, the flux density variability of PKS~0003$-$066 is shown
in Fig. 11 for 8.6 GHz and Fig. 12 for 2.3 GHz, using the all VLBI data analysed, 
decomposing the source into its different pc-scale components. 

An interpretation of the flux density monitoring observations 
such as the Australia Telescope Compact Array (ATCA) (Tingay et al. 2003) 
and the University of Michigan 26m (UMRAO) database
(see e.g. Aller et al. 1985 for description of the monitoring program) 
will appear elsewhere (Horiuchi \& Tingay 2006, in preparation).

\section{Discussion}

\subsection{Proper motions and variability of the jet components}

From their analysis of 5 epochs of data over 5.5 years (from July 1995 to January 2001) 
at 15 GHz, Kellermann et al. (2004) found no siginificant
proper motions for any pc-scale jet components in PKS~0003$-$066.  
In contrast, from an examination of the montages in Figs. 6 and 7, 
substantial evolution of the source structure is
apparent when considering a large amount of data over a ten year period.
We fit the radial distances from the core versus time for each component 
of the 8 GHz images with the linear motion that minimizes the ${\chi}^2$ (Fig. 9 (a)). 
The uncertainty of the proper motion, defined by a significance 
of $\ge 3 \sigma$, is estimated using  the ${\chi}^2$ 
statistics that includes both scattering of the data points around the proper 
motion-model and uncertainties of individual observations 
(one quarter of the beam sizes are adopted). 
We show that the inner jet components, 
denoted C2 and C4 (see Fig. 9 (a)), have proper motions 
of 0.58$\pm$0.02 and 1.35$\pm$0.05 mas/yr, corresponding to
apparent speeds of 9.3$\pm$0.3 $c$ and 21.6$\pm$0.8 $c$.  
These inner components are much faster than the outer component, C1,
which is moving at 0.12$\pm$0.01 mas/yr or approximately 2.0$\pm$0.2 $c$
(Table 8). 

In Fig. 9 (b), the 15 GHz model fit components by Kellermann et al. (2004)
based on their image analysis (no error estimates on their component position
were provided) are superposed on the 8.6 GHz fits for the linear motions
together with the 2.3 GHz model components. The 2.3 GHz components
follow the similar trend of proper motions as seen at 8.6 GHz. 
Although there is a hint
of shifts within one milliarcsecond or so between the paths at 
2.3 and 8.6 GHz, the positional acuracy of 2.3 GHz data, 
typically about 0.5 mas, is comparable to this.
 
The reason why Kellermann et al. (2004) didn't identify such high 
proper motions may be attributed to the misidentification of the components
due to the large gaps between observations. 
We have data points almost 
every two months from 1997 to 1998, hence we clearly see the fast proper
motion of C2 continuously over that period.  
The fact that the results of proper motion studies of core-dominated radio 
sources can be so affected by the sampling frequency and overall period of 
monitoring is well known (e.g. G\'{o}mez et al. 2001).
Moreover, Fig. 9 (a),(b) indicates that our data point for C2 at the epoch of 
16 January 2002 (MJD 52291) is consistent with those for the component C
of Kellermann et al. (2004) around that epoch. 
Also our images are at a lower frequency, hence steep spectrum components such 
as C1, C2 and C4 are brighter (see Fig 12(b)) and easier to identify in spite of the
lower spatial resolution.
In contrast, at 15 GHz some jet components are resolved and are weaker, 
making a consistent identification of the overall jet structure more difficult.
Note also that Kellermann et al. (2004) derived positions 
in the image plane while we derived positions in the UV plane. This may lead 
to some differences in the derived positions for the components.

\subsubsection{Kinematics of the fast jet components}

Adopting the relation between the intrinsic jet speed $\beta = v/c <$ 1, 
the apparent speed $\beta_{{\rm app}}$, and the viewing angle of the 
jet to our line of sight $\theta$  
$$ \beta_{{\rm app}} = \frac{\beta \sin\theta}{1-\beta \cos\theta}, $$
we find that the jet inclination that maximises the apparent speed is 
$\theta_{{\rm max}} = 5.3^\circ$, using the fastest apparent motion, 
$\beta_{{\rm app}} = 21.6$, of C4 or $12.3^\circ$ for the $9.3c$ apparent motion of C2. 
One can then derive a minimum Lorentz factor, $\gamma_{{\rm min}} = 21.6$ for C4, 
or 9.35 for C2, 
which corresponds to a Doppler factor of 
$\delta_{\rm min} = 
\gamma_{\rm min}^{-1}(1-\beta_{\rm min}\cos\theta_{\rm max})^{-1}$ = 8.63
for C4 and 3.74 for C2,
where $\beta_{{\rm min}} = \sqrt{1-\gamma_{{\rm min}}^{-2}}$ = 0.9989 for C4 
or 0.9942 for C2.    
For smaller viewing angles the Doppler factor approaches $\delta \sim 2\gamma$ 
as $\theta \sim 0$, which yields $\delta_{\rm max} = 43.2$ for C4 and 18.7 for C2. 
Although Kellermann et al. (2004) didn't measure superluminal 
motions in PKS~0003$-$066, in thier sample of 110 sources for the VLBA 2cm suevey
they measued superluminal motions ranging up to $\beta_{app} \sim 25$.
Our results suggest that the superluminal motion of PKS 0003-066 we measure 
corresponds to the middle to high end of the distribution of superluminal motions 
in the sample of Kellermann et al. (2004, Fig.10). 

\subsubsection{The interaction between the moving and quasi-stationary components}

Also clear from the montage of images and Fig. 9 is that the fast components 
C2 and C4 catch up to the slower component C1 and interact strongly near epoch 1999.  
As this interaction occurs, the 
C1 region clearly gets brighter.  The component C3, closest to the core, appears stationary.

Coexistence of faster and slower or quasi-stationary components are 
seen in many sources such as 
4C 39.25 (Alberdi et al. 2000), 
Centaurus~A (Tingay et al. 2001), 3C 120 (G\'{o}mez et al. 2001), 3C~279 (Jorstad et el. 
2004) and 1803+784 (Britzen et al. 2005), which may be 
explained within relativistic 
time-dependent hydrodynamic models of ``trailing shocks" (Agudo et al. 2001).
In this scenario it is expected that the motion of inner components 
is slower than that of the outer primary component, as clearly seen in 3C~120 
(G\'{o}mez et al. 2001) and Centaurus~A (Tingay et al. 2001).
Our analysis of PKS~0003$-$066 suggests that C3 may be trailing shock for C2. 

Another explanation of the stationary features could be as bends of the jet 
trajectory in a plane different to the observers plane, as suggested for
4C~39.25 by Alberdi et al. (2000). 
In PKS~0003$-$066 the component C1 has a non-radial motion, from NE to SW, while
brightning in total flux density. The component C3 is stationary in the distance from 
the core but not stationary in the direction perpendicular to the radial direction.

The interaction between C1 and C2 may be also evident  in the variability 
of the polarization profile for C1. 
The VLBA polarization images at 15 GHz from 2003 February to 2005 June
(Lister \& Homan 2005, MOJAVE database website) show that the component C1 is 
significantly polarized, $\sim$20\% average and up 
to 60\% around the outer edge of the component, suggesting compression and 
enhancement of the local magnetic field at the interface with the ambient media. 
Such high fractional polarization 
is not seen for the C1 component in the 5 GHz global VLBI image from 1995 May 
(Gabuzda, Pushkarev \& Cawthorne 2000), with only an upper limit of $<$11\%, 
possibly because the component is not compressed enough at that stage. 

The brightning of a jet caused by internal shocks with a slightly 
varying jet speed is seen also in the well studied galactic X-ray binary jet 
SS~433. Migliari, Fender \& M\'{e}ndez (2002) show evidence for a hot continuum and 
Doppler-shifted iron emission lines from spatially resolved regions, 
suggesting that {\it in situ} reheating of the jet component takes place in a 
flow that moves with relativistic bulk velocity more than 100 days after 
launch from the binary core of the SS~433 jet. 


\subsubsection{Relation to large-scale structure}

As discussed earlier, the pc-scale VLBI jet kinematics possibly
supports jet bending. 
An interesting question is whether the structure on kpc-scales as seen 
with the VLA would be related to the pc-scale structure with an 
$\sim$90$^\circ$ misalignment, as commonly 
seen in many compact core-dominant sources (e.g. Conway \& Murphy 1993). 
In a report of a VLA calibrator position survey, Perley (1982) noted a single secondary 
component of PKS~0003$-$066 located at $r=1.7$\arcsec~ (8.5 kpc), p.a=30$^\circ$, but to 
our knowledge no kpc-scale image for this source has ever been published. 
We made two VLA images from archived data at 4.9 and 8.5 GHz 
(Figs. 13(a) and (b)). We clearly detect the secondary component as pointed out by 
Perley (1982), although it appears very weak, approximately 0.3 percent
of the flux density from the unresolved core. 
The inner VLBI jet ($<$ 0.8 mas) has an initial position angle centered at 7$^\circ$ with 
an oscillation of $\pm$15$^\circ$. On scales of 2-6 mas, the position angle 
is close to $\sim$90$^\circ$. On arcsec scales, the jet appears close to 
0$^\circ$ once again. 
A natural explanation of the extreme curvature of the jet is that 
we are looking at a part of a spiral-like trajectory, 
its appearance affected by projection onto the sky plane.
An example of such orthogonal bending in a kpc-scale jet of a BL Lac object 
can be seen in 1803$+$784 as discussed by Tateyama et el. (2002), modeling a
helical jet structure. 


\section{Conclusion}

The main results of our multi-epoch VLBI image analysis of the BL Lac PKS~0003$-$066
are as follows:

1. In contrast to previous studies of this source, VLBI components of the jet 
are highly variable, with proper motions of two components, C2 and C4, found
to be as high as 0.6 mas/yr and 1.4 mas/yr ($\sim$ 9 $c$ and 22 $c$) respectively.
The jet component C1 is moving very slowly with $\sim$0.1 mas/yr
or $\sim$2.0 $c$. As the inner component C2 approaches C1, the flux density of C1 
increases 3 times, indicating an interaction between the components.


2. A new disk-based recording system and software correlator for VLBI were verified
by comparing imaging results to other established recorder/correlator systems.

\acknowledgments
This work is supported by the Australian Federal Governments
Major National Research Facility (MNRF) program.
The Australia Telescope Compact Array is funded by the 
Australian Commonwealth Government for operation as a national 
facility managed by CSIRO.  
The VLBA is an instrument of the National Radio Astronomical Observatory, 
which is a facility of the National Science Foundation operated under 
cooperative agreement by Associated Universities, Inc. 
This research has made use of the United States Naval Observatory (USNO) 
Radio Reference Frame Image Database (RRFID).
This research has made use of data from the University of Michigan Radio 
Astronomy Observatory, which is supported by the National Science 
Foundation and by funds from the University of Michigan. 
This research has made use of the NASA/IPAC Extragalactic Database (NED) 
which is operated by the Jet Propulsion Laboratory, California Institute 
of Technology, under contract with the National Aeronautics and Space Administration.

\clearpage



\begin{figure}
\plotone{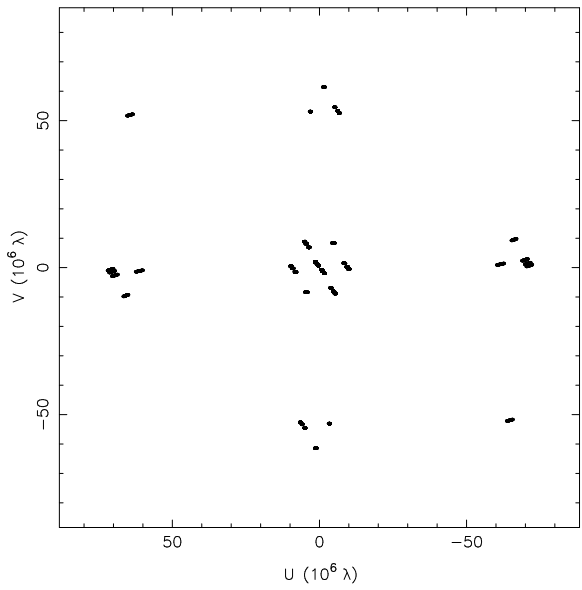}
\caption{
($u,v$) coverage of the global VLBI observations 
of 2004 April 18 at 2.3 GHz.
}
\end{figure}

\begin{figure}
\plotone{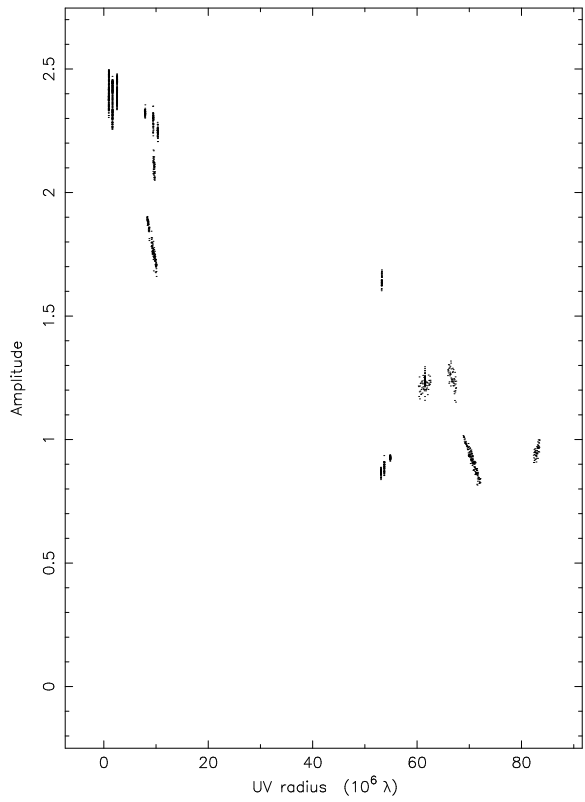}
\caption{
Visibility amplitude in Jy versus ($u,v$) distance, for the data shown in Figure 1.\label{fig1}
}
\end{figure}

\clearpage


\begin{figure}
\plotone{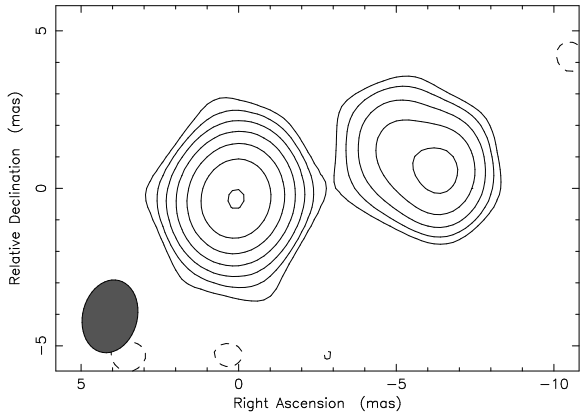}
\caption{Image obtained from the global VLBI observations of 2004 April 18
at 2.3 GHz.
The FWHM of the restored beam is 2.34 $\times$ 1.73 mas at position angle 
-13.9$^\circ$,
as indicated by the filled ellipse in the bottom left hand corner.
Contours are drawn at
-1.5, 1.5, 3, 6, 12, 24, 48, and 92\% of 1.31 Jy/beam,
the peak flux density in the map.\label{fig2}}
\end{figure}




\begin{figure}
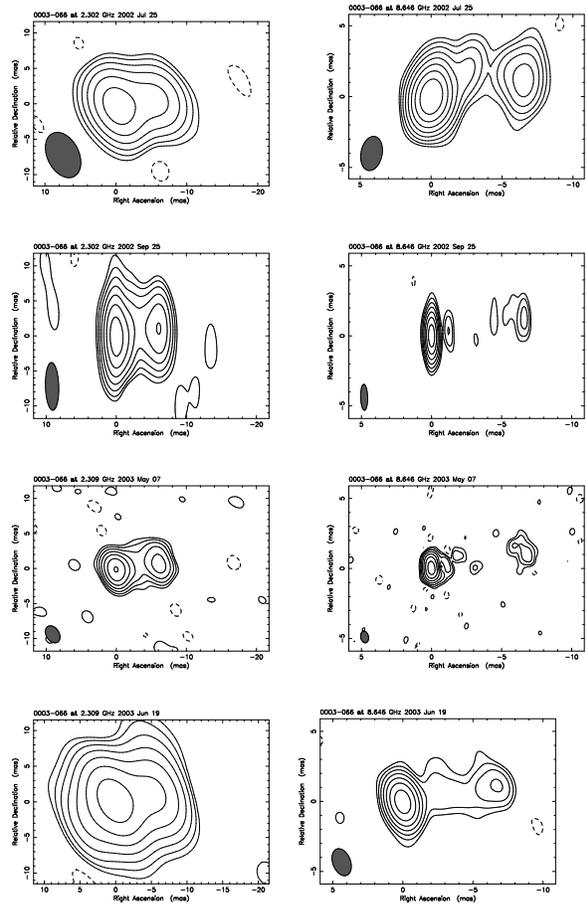

\plottwo{f3a.ps}{f3b.ps}
\plottwo{f3c.ps}{f3d.ps}
\plottwo{f3e.ps}{f3f.ps}
\plottwo{f3g.ps}{f3h.ps}

\caption{VLBA observations at 2.3 and 8.6 GHz.}
\end{figure}

\begin{figure}
\plottwo{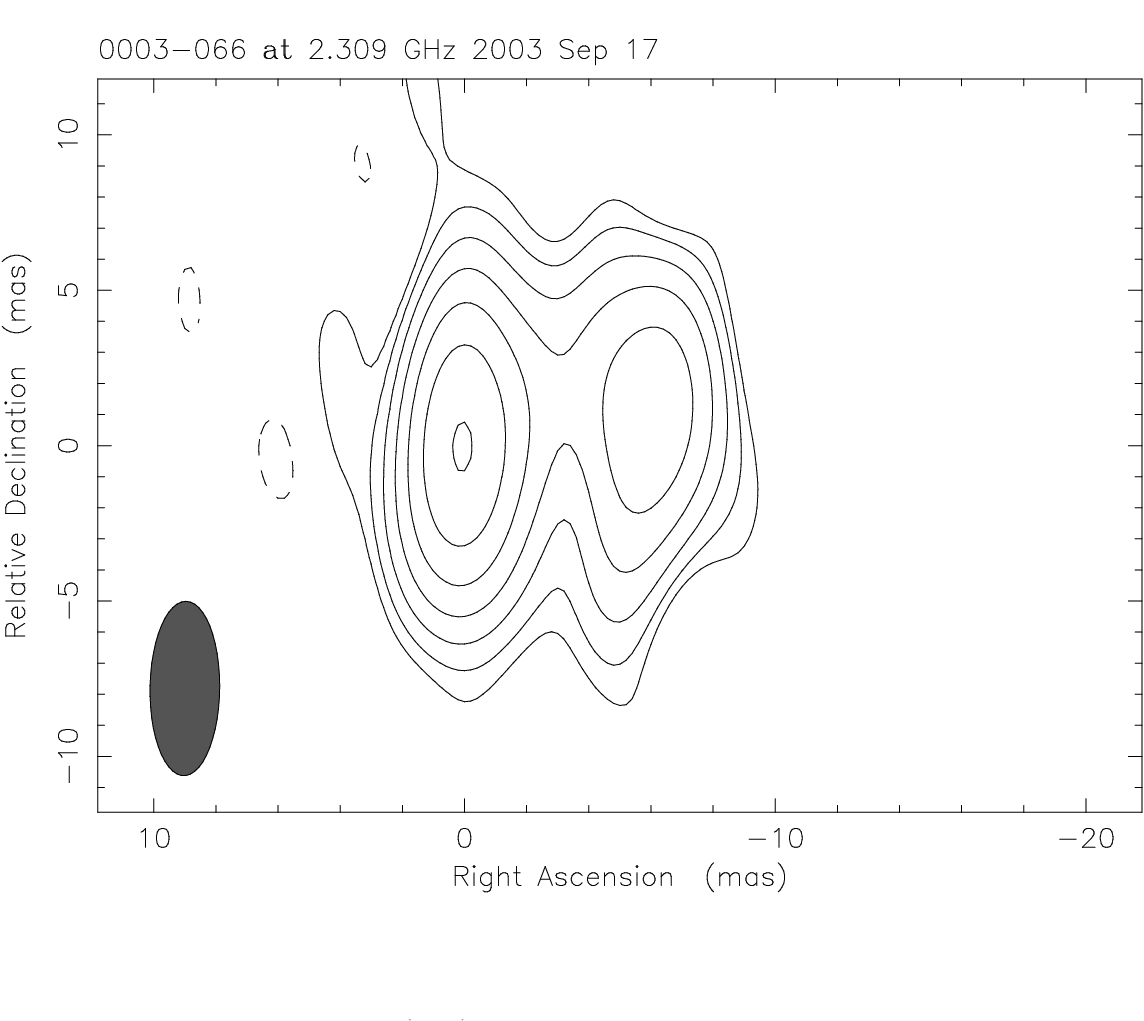}{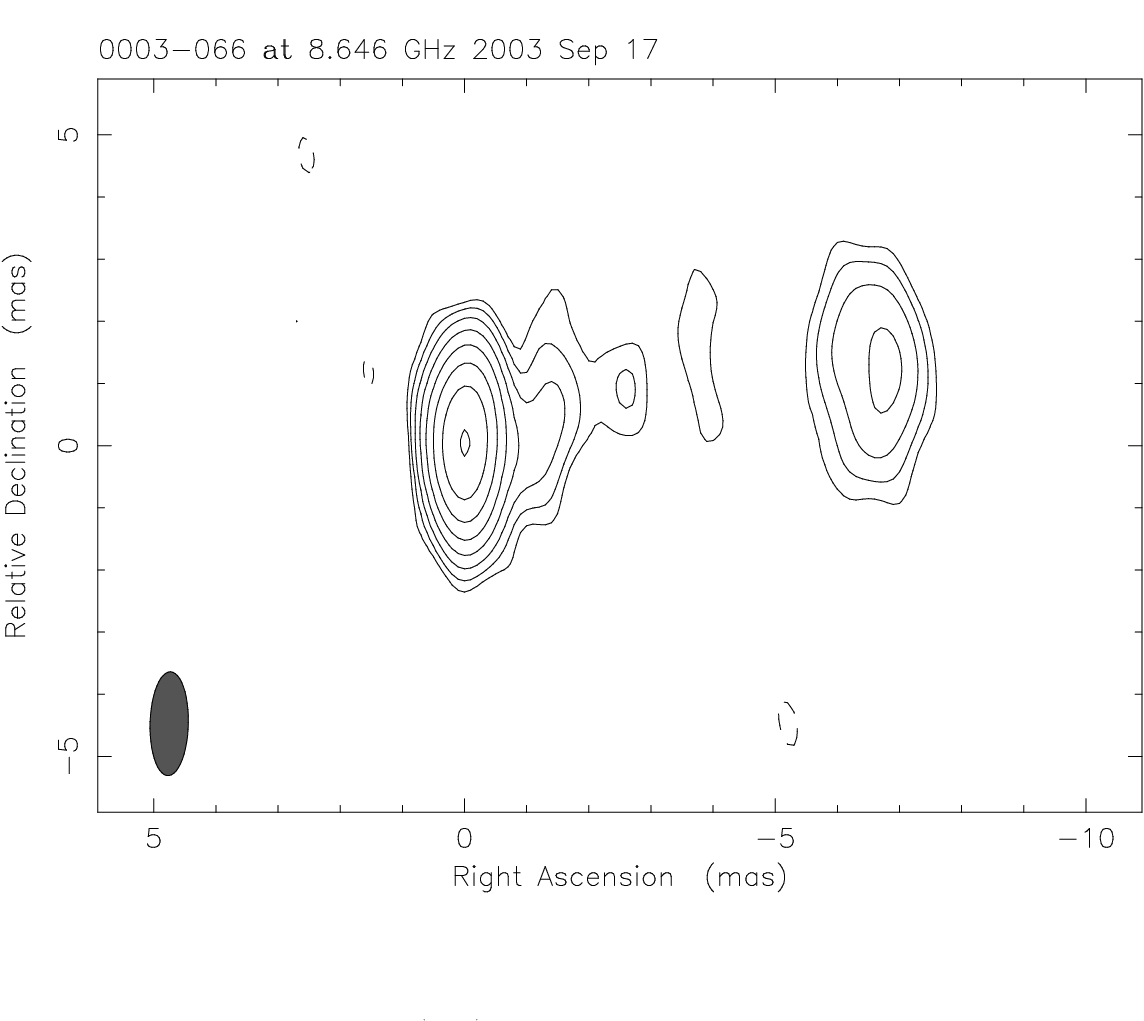}
\plottwo{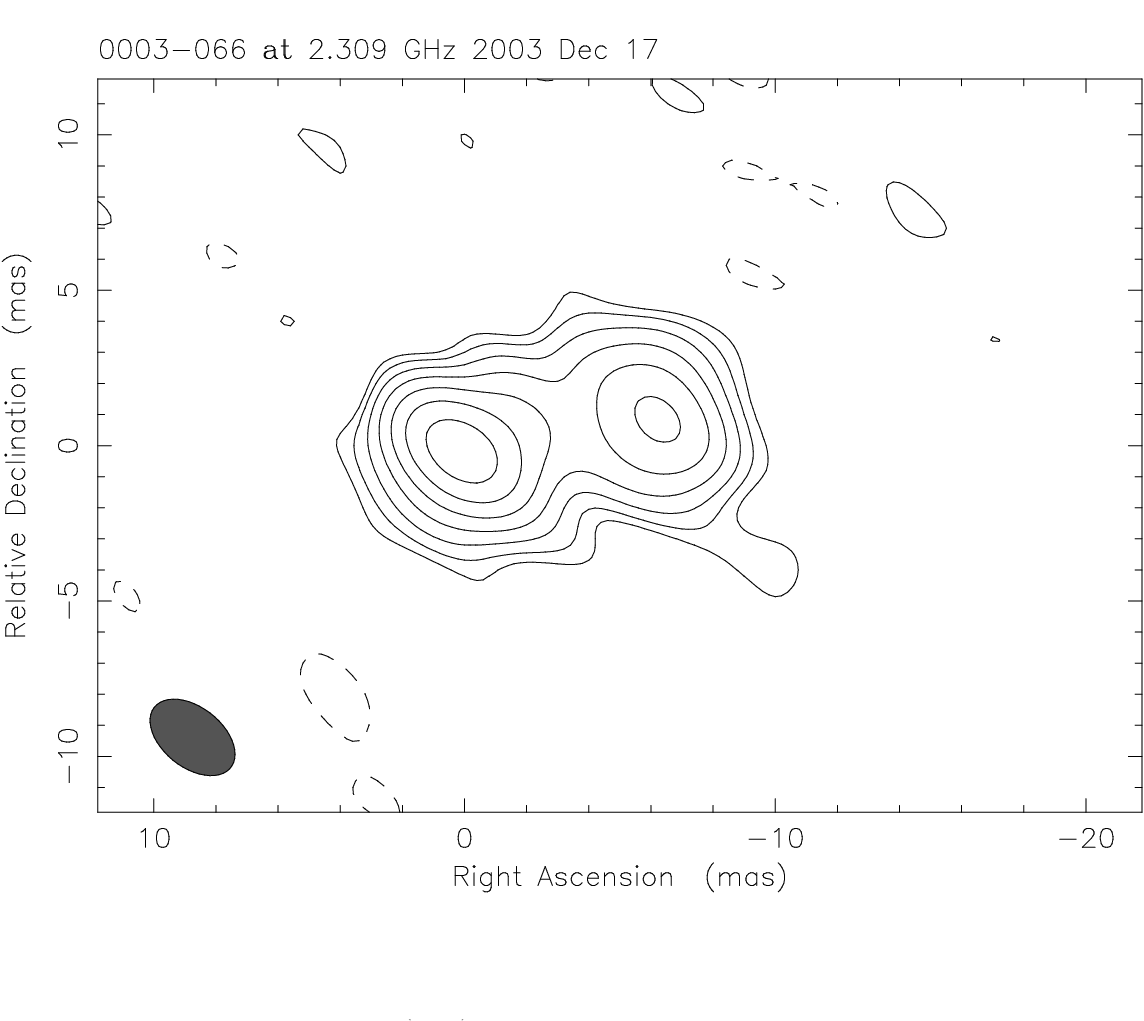}{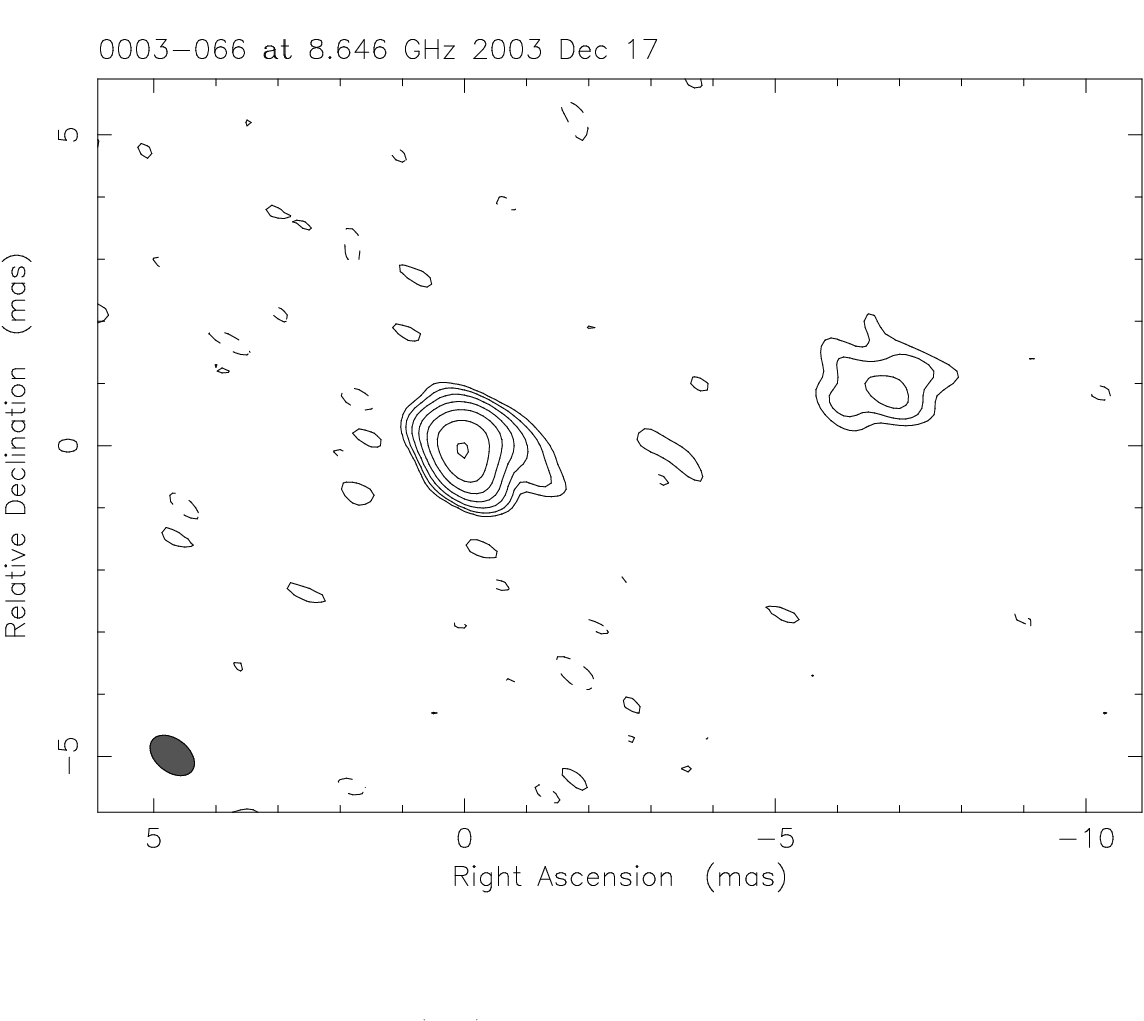}
\plottwo{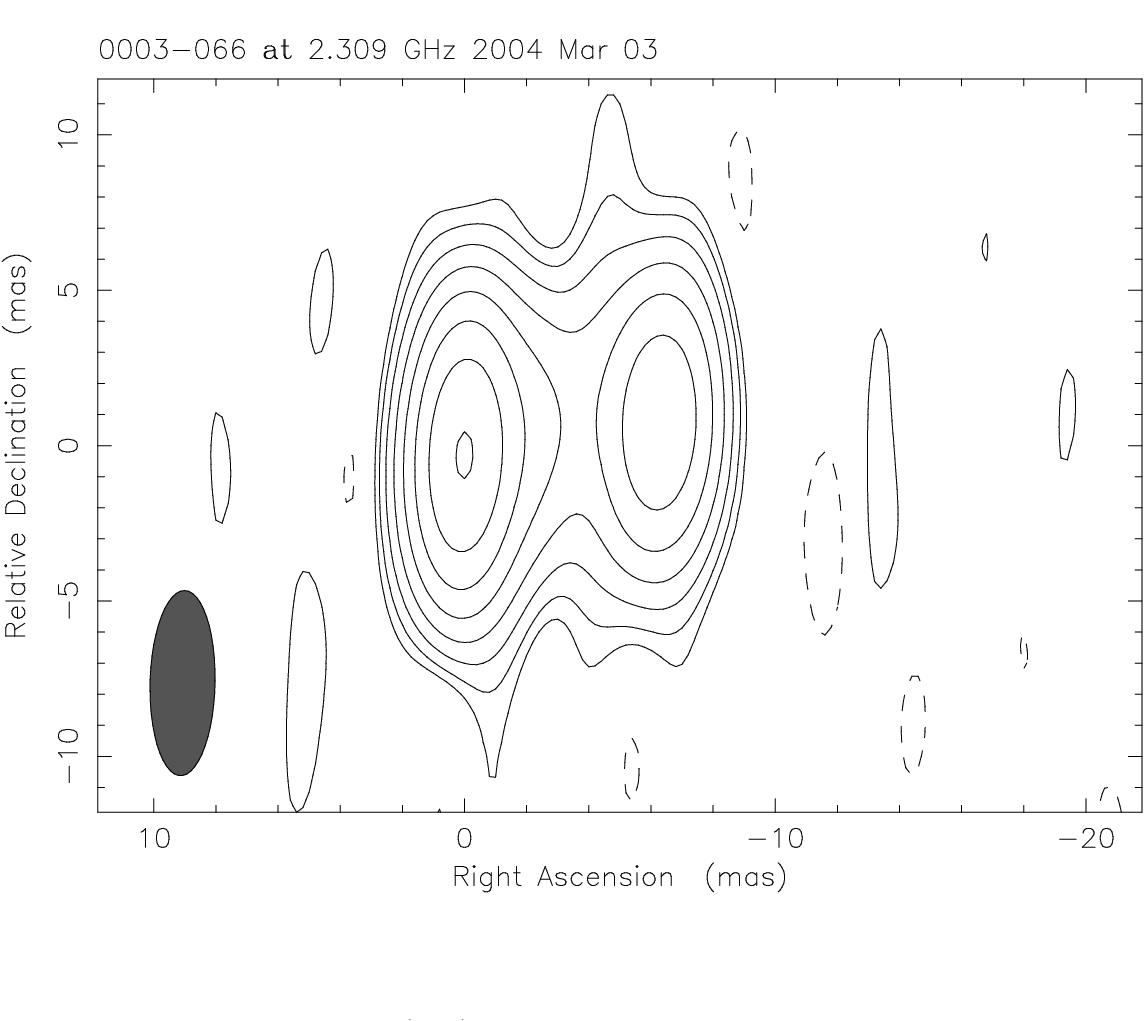}{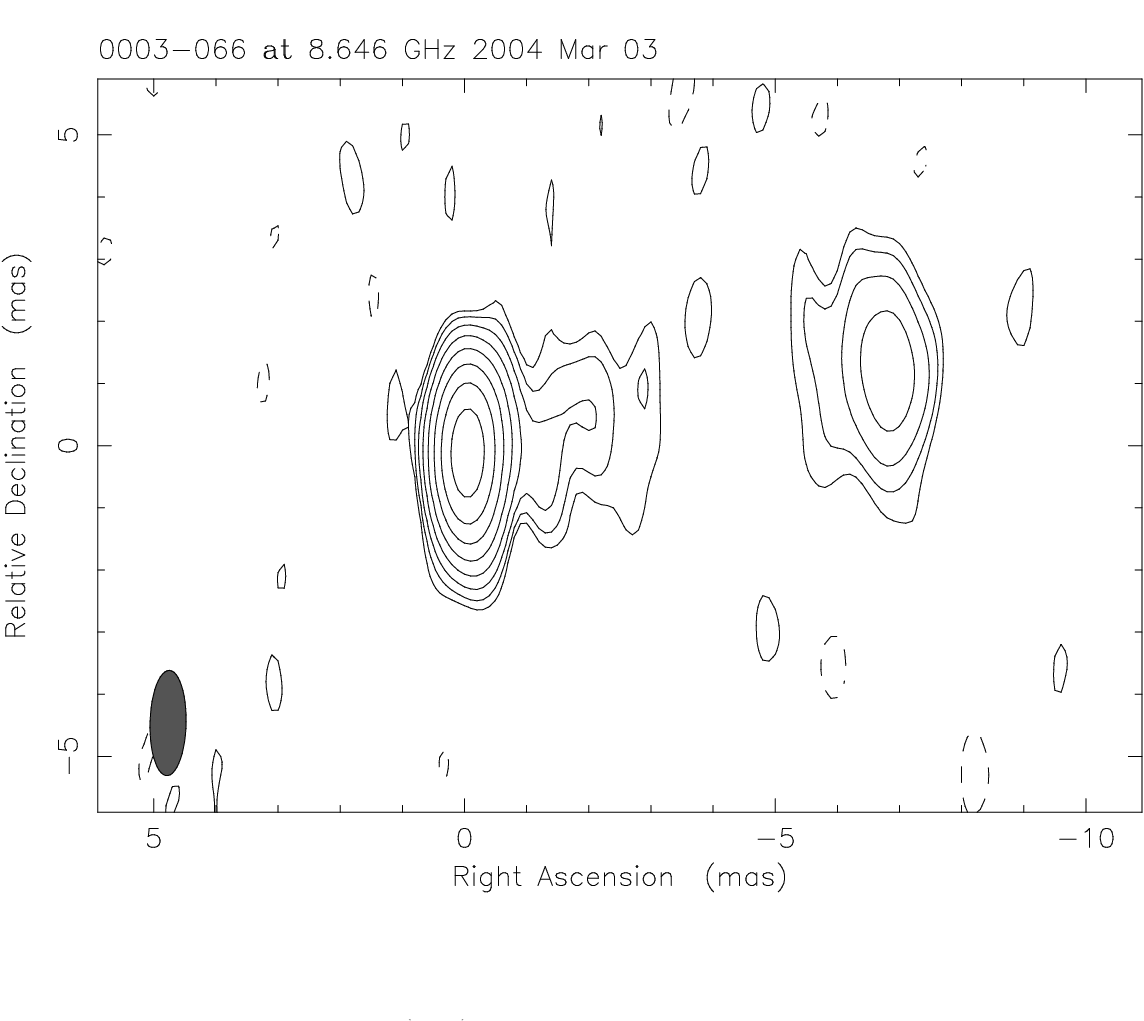}
\plottwo{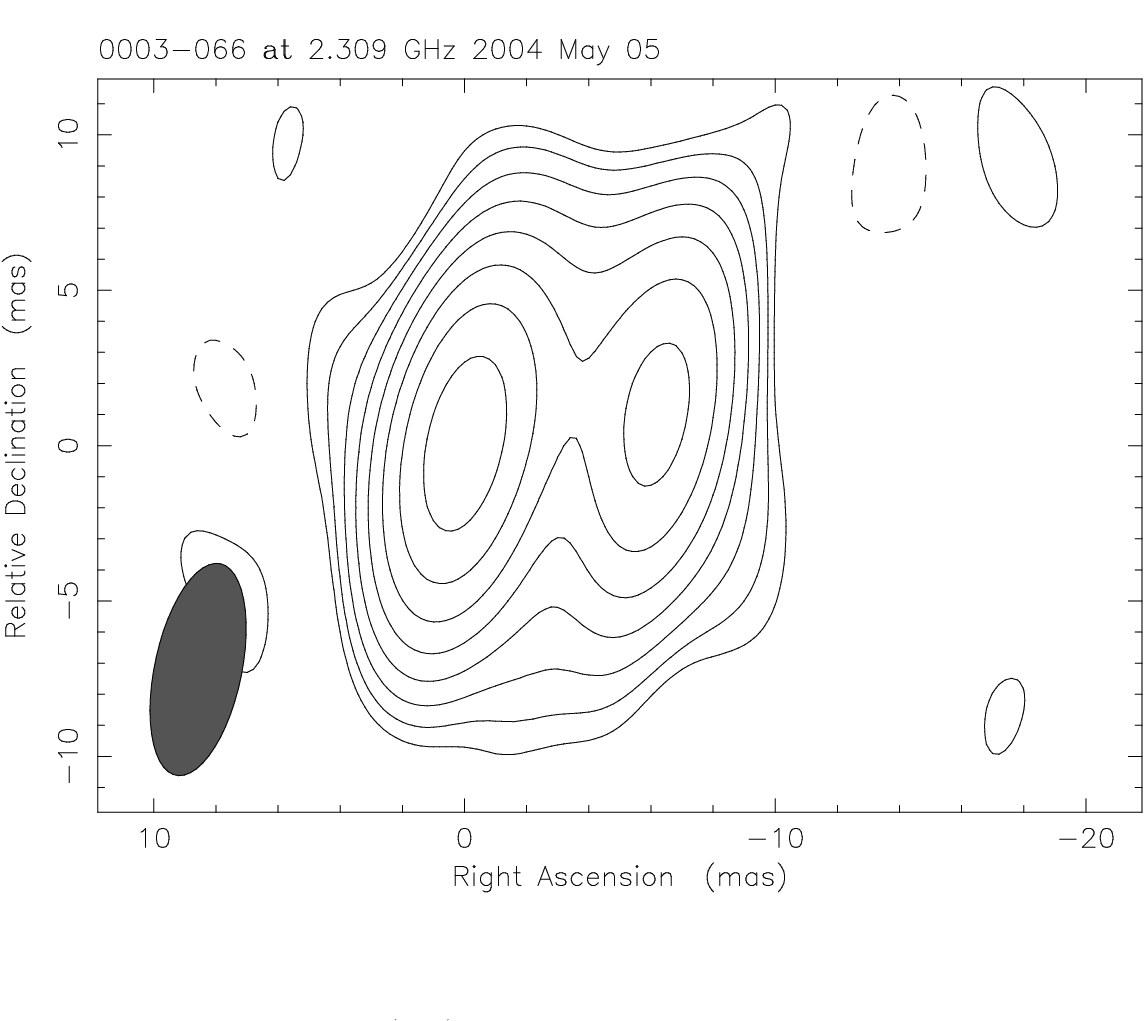}{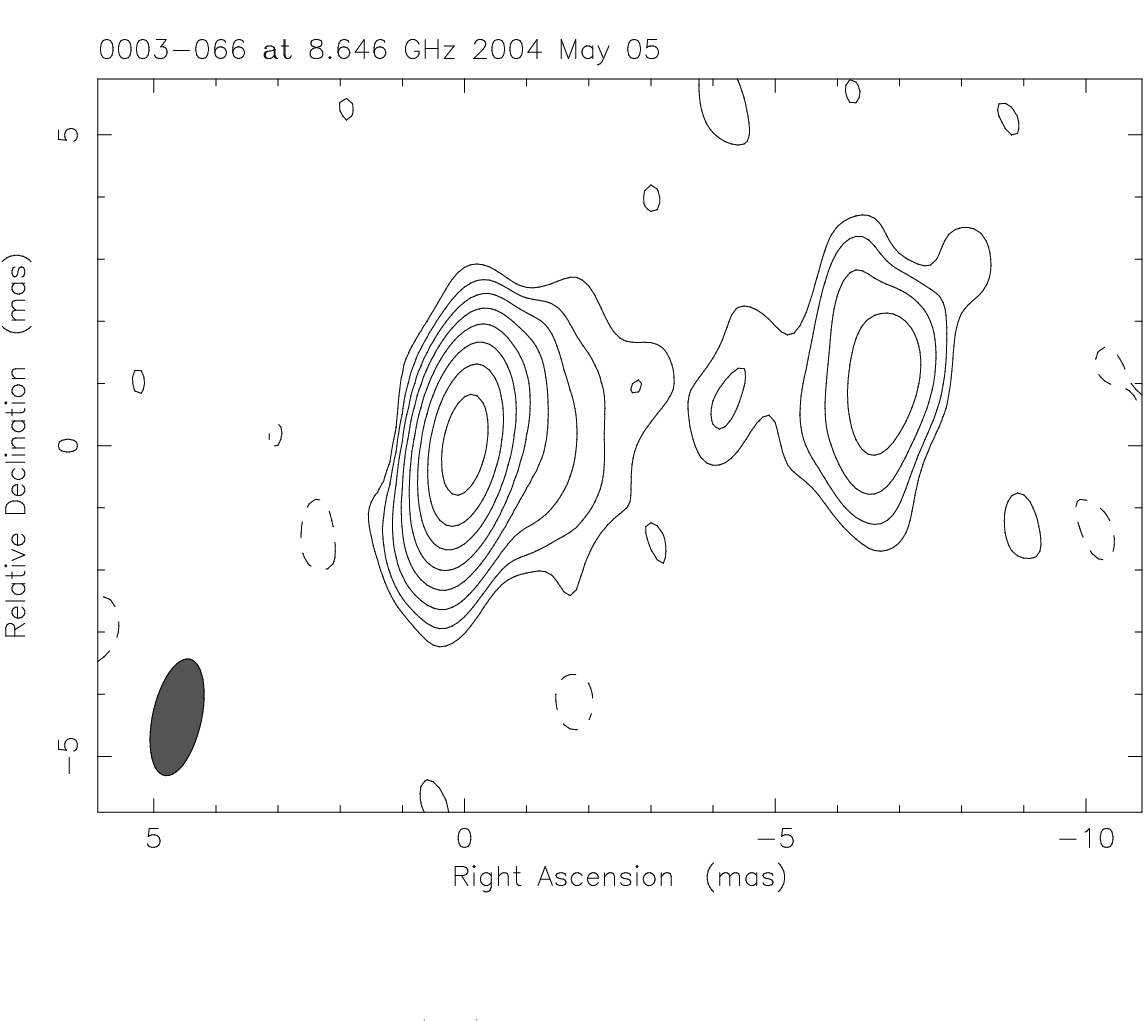}
\caption{VLBA observations - cont.}
\end{figure}

\begin{figure}
\plottwo{f5a.ps}{f5b.ps}
\plottwo{f5c.ps}{f5d.ps}
\plottwo{f5e.ps}{f5f.ps}
\plottwo{f5g.ps}{f5h.ps}
\caption{VLBA observations - cont.}
\end{figure}

\begin{figure}
\epsscale{.56}
\plotone{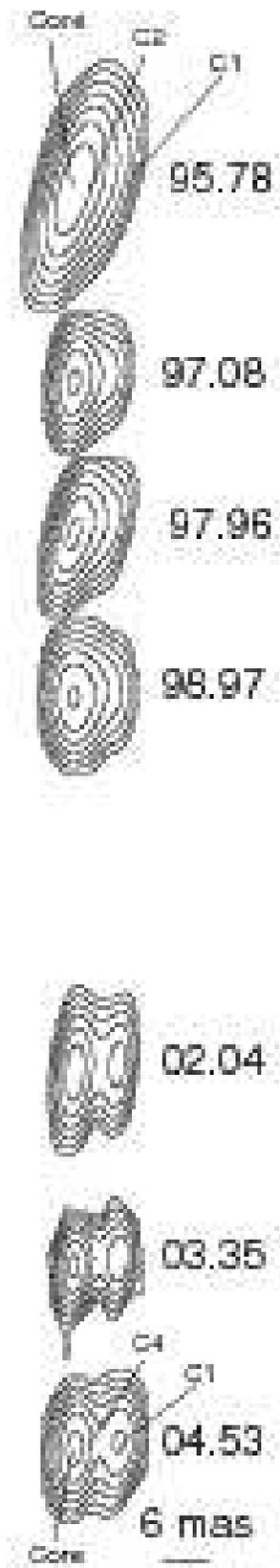}
\caption{Time series mosaic of a selection of 2.3 GHz VLBI images of PKS~0003$-$066.\label{fig3a}}
\end{figure}

\begin{figure}
\epsscale{.40}
\plotone{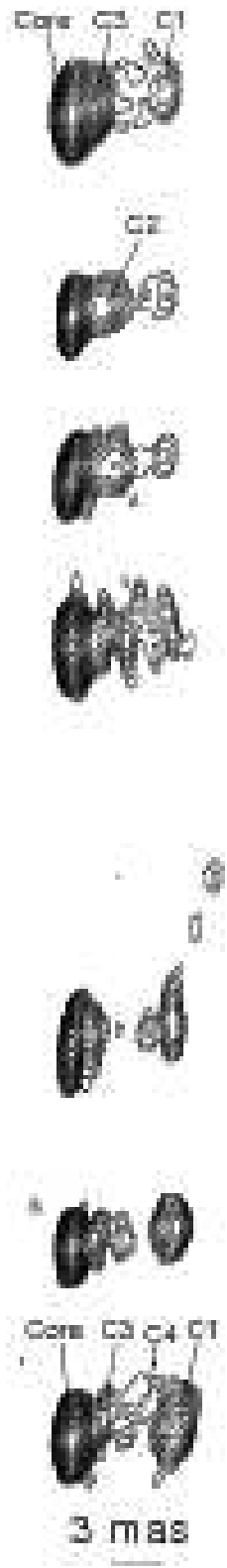}
\caption{Time series mosaic of a selection of 8.6 GHz VLBI images of PKS~0003$-$066
(observing epochs are the same as 2.3 GHz).\label{fig3b}}
\end{figure}

\begin{figure}
\epsscale{.90}
\plotone{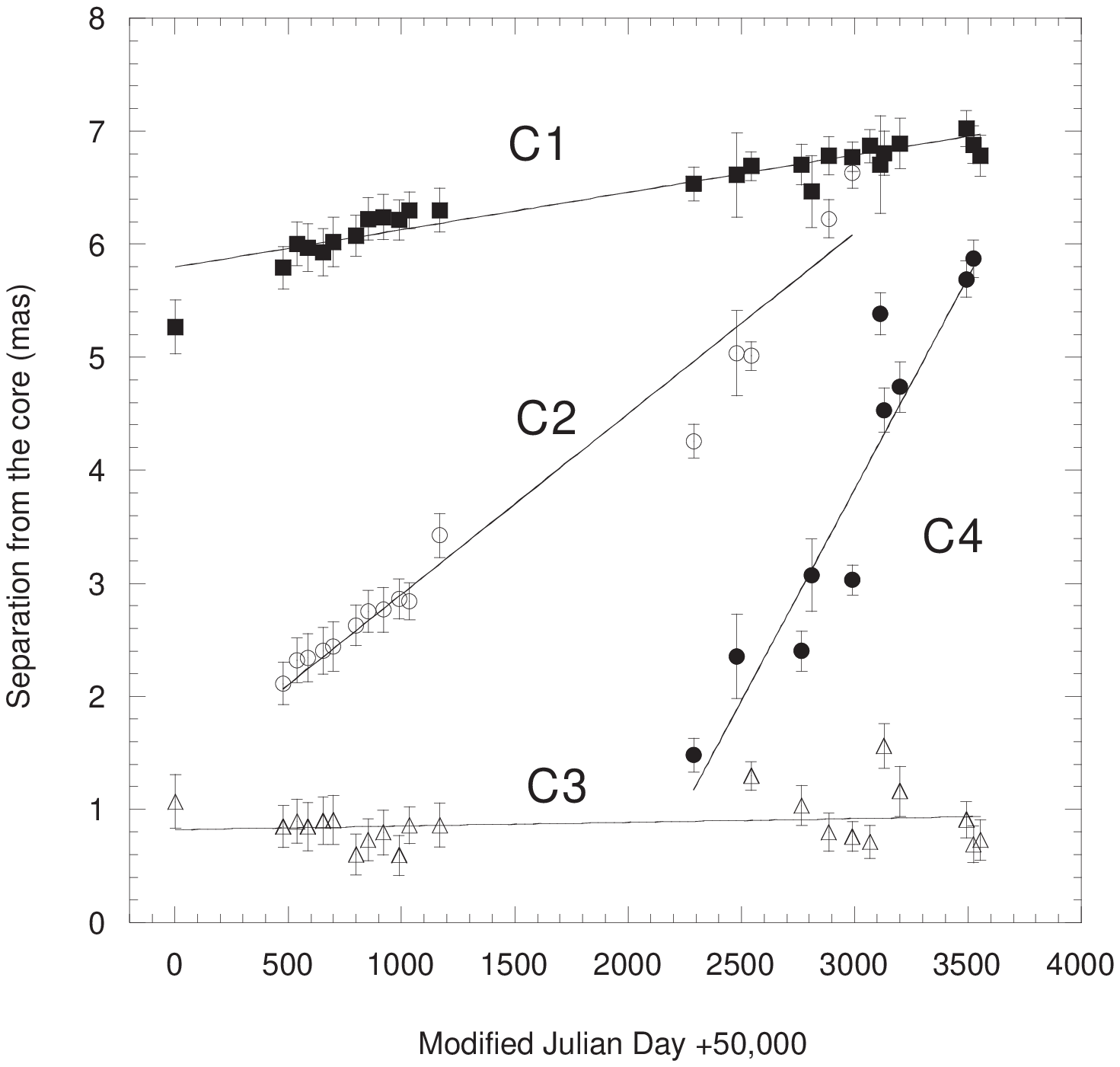}
\plotone{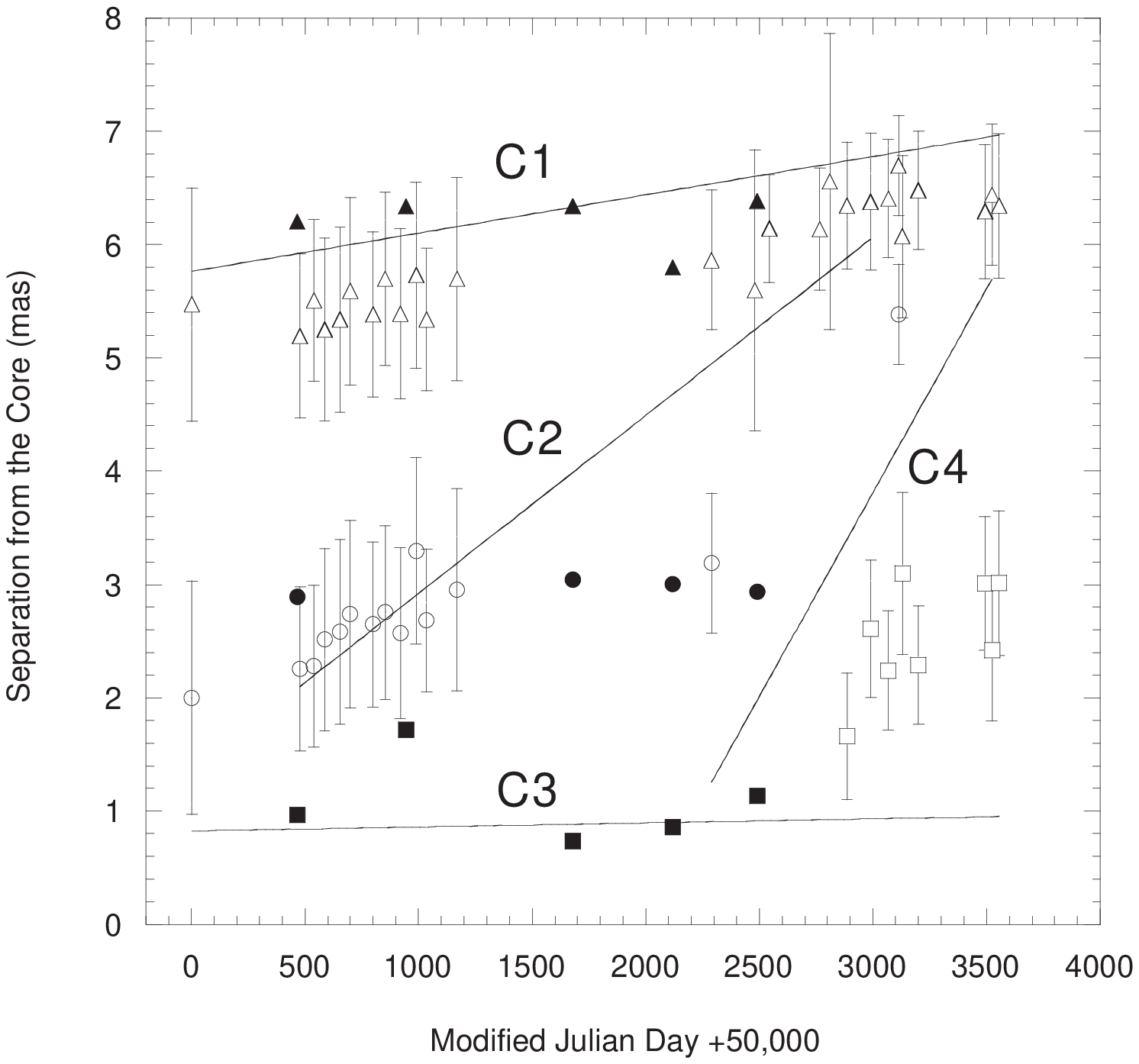}
\caption{(a) (top) Angular distance from the core of Gaussian components (C1, C2, C3 and C4) 
as a function of time (date from 50,000 Modified Julian Date, 
i.e. 1995 October 10) for observations with the VLBA at 8.6 GHz and 
the global VLBI observations of 2004 April 18
 at 2.3 GHz (MJD 53,114).
The lines are the linear least-square fits to the data. 
(b) (bottom)  The same 8.6 GHz fitted lines as the above but, for comparison, 
with 2.3 GHz components  (triangles C1, circles C2, and square C4) 
and 15 GHz components shown by Kellermann et al. (2004)
(filled squares component B, filled circles component C, and filled triangles component D).
}
\end{figure}

\begin{figure}
\epsscale{.90}
\plotone{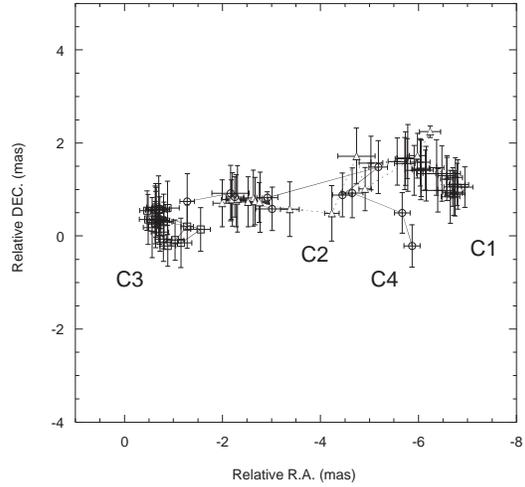}
\caption{Position relative to the core of each component C1, C2, C3, C4
at all epochs derived from the observations with the VLBA at 8.6 GHz.
Note that positional errors are estimated to be typically 0.2 and 0.5 mas for 
R.A. and Dec. respectively at 8.6 GHz taking one quarter of the beam 
sizes.}
\end{figure}

\begin{figure}
\epsscale{1.20}
\epsscale{.90}
\plotone{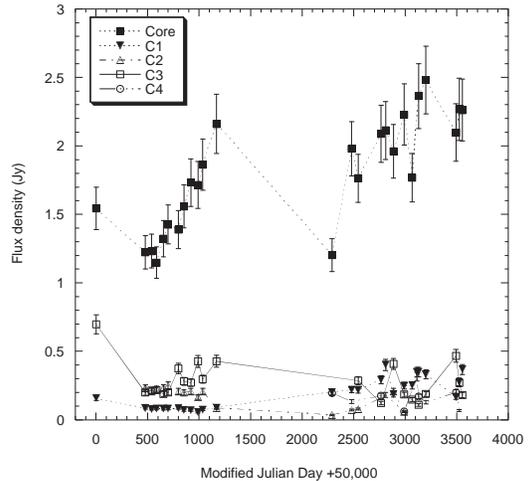}
\caption{Flux density variation for each VLBA 8.6 GHz component 
 as a function of time (date from 50,000 Modified Julian 
Date, i.e. 1995 October 10). 
The error bars on the flux density points
are estimated at 10 percent of the flux density at each component. 
}
\end{figure}

\begin{figure}
\epsscale{.90}
\plotone{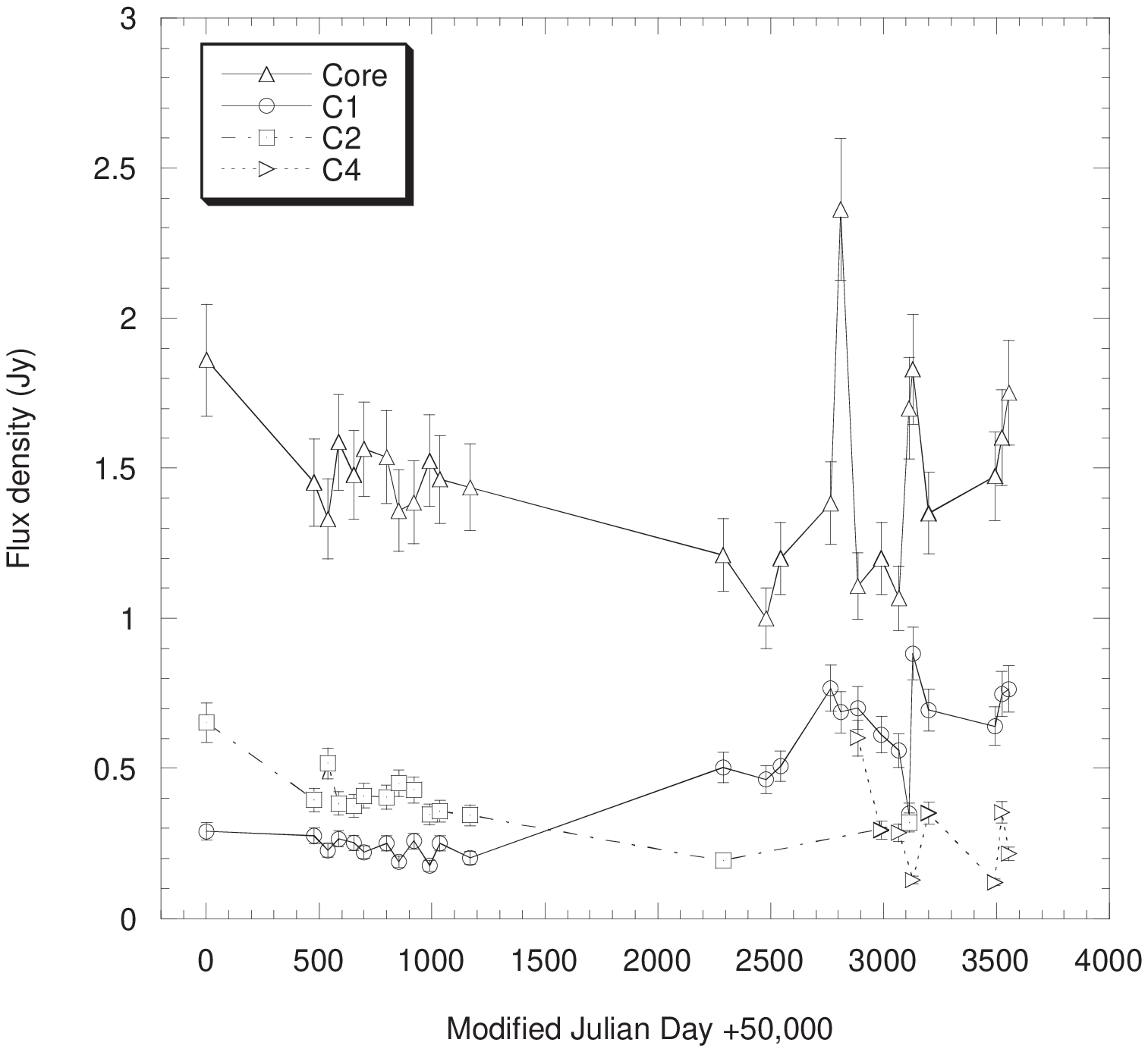}
\plotone{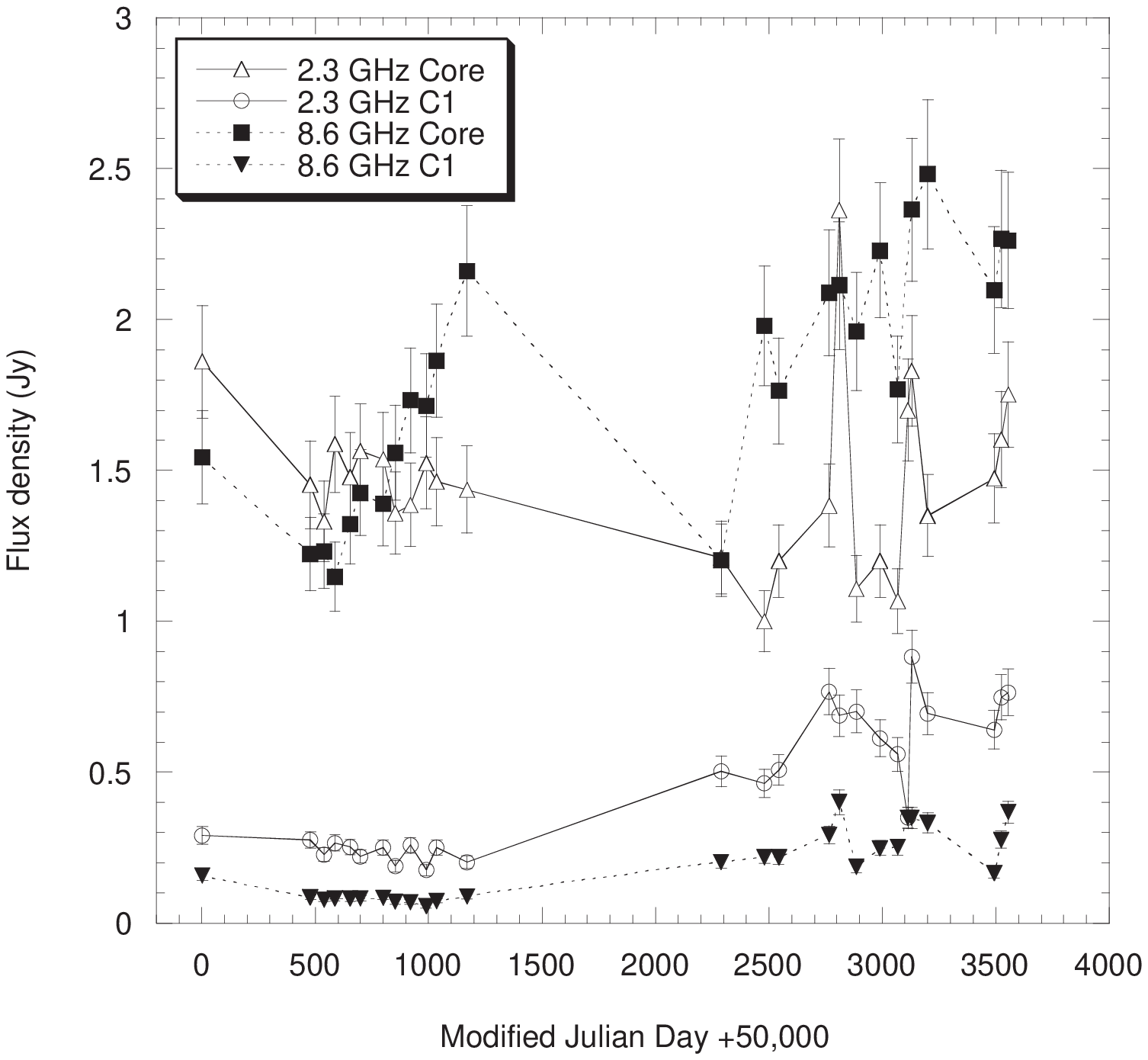}
\caption{(a) (top) The flux density of each VLBI component at 2.3 GHz.
(b) (bottom) The total flux density of the VLBI core components and outer 
most components C1 at 2.3 GHz and 8.6 GHz.  
}
\end{figure}

\begin{figure}
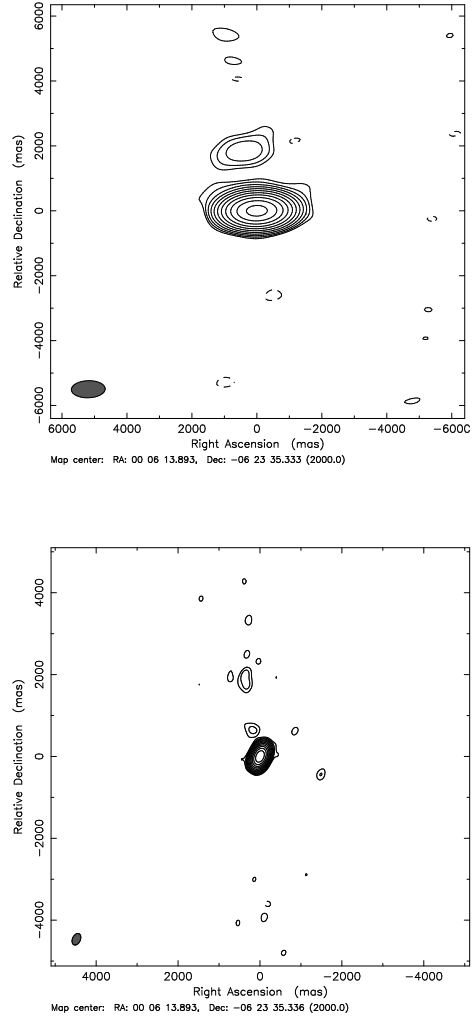

\epsscale{0.8}
\plotone{f13a.ps}
\plotone{f13b.ps}
\caption{Kiloparsec scale structure of PKS~0003$-$066 imaged from VLA
archive data: 
(a) (top) at 4.86 GHz on 1997 January 25, 
with the FWHM of the restored beam 1.04 $\times$ 5.26 (arcsec) at P.A. -88.3$^\circ$, and 
with contours at -0.075, 0.075,
0.15, 0.3, 0.6, 1.2, 2.4, 4.8, 9.6, 19.2, 38.4, 76.8\% of 2.46 Jy/beam, the peak flux 
of the map,    
(b) (bottom) at 8.46 GHz on 2002 February 21,
with the FWHM of the restored beam 2.99 $\times$ 1.94 (arcsec) at P.A. -24.6$^\circ$, and 
with contours at -0.05, 0.05,
0.1, 0.2, 0.4, 0.8, 1.6, 3.2, 6.4, 12.8, 25.6, 51.2\% of 2.76 Jy/beam, the peak flux 
of the map. Both observations were made with the VLA A-configuration.}
\end{figure}







\clearpage
\begin{table}
\begin{center}
\caption{Antennas participating in 2004 April 18 global VLBI observations at 2.3 GHz}
\begin{tabular}{lccccl}
\tableline\tableline
Antenna & Organisation\tablenotemark{a} & D (m) & SEFD\tablenotemark{b} (Jy) & Recording system \\
\tableline
ATCA &	ATNF &	5 $\times$ 22 & 90 &	LBADR\\
Mopra & ATNF & 22 & 400 & LBADR\\
Parkes & ATNF & 64 & 45  & LBADR\\
Hobart & U.Tas.	&	26	& 650	&	Mark5\\
Ceduna	&	U.Tas.	&	30	& 350	&	LBADR\\
Hartebeesthoek &	Hart.RAO &	26 &	390	&	Mark5\\
Kashima	&	NICT	&	34 &	400	&	K5\\

\tableline
\end{tabular}
\tablenotetext{a}{ATNF is Australian Telescope National Facility. U.Tas. is 
University of Tasmania (Australia). Hart.RAO is Hartebeesthoek Radio Astronomy
Observatory (South Africa). NICT is National Institute of Information and 
Communications Technology (Japan)}
\tablenotetext{b}{System Equivalent Flux Density}
\end{center}
\end{table}

\clearpage
\begin{table}
\begin{center}
\caption{Gaussian model of the image from global VLBI observations of 2004 April 18 at 2.3 GHz
}
\begin{tabular}{lcccccl}
\tableline\tableline
Component & $S$(Jy)\tablenotemark{a} & $r$(mas)\tablenotemark{b} & PA(deg.)\tablenotemark{b}
 & $a$ (mas)\tablenotemark{c} & $b/a$\tablenotemark{c} & $\psi$(deg.)\tablenotemark{d} \\
\tableline
Core & 1.70 & - & - & 1.24 & 0.66 & 74.1 \\
C1 & 0.35 & 6.70 & -82.8 & 0.70 & 0.00 & 43.0 \\
C4 & 0.32 & 5.38 & -74.1 & 1.68 & 0.36 & 49.3 \\
\tableline
\end{tabular}
\tablenotetext{a}{Flux density in Janskys}
\tablenotetext{b}{r and PA are the polar coordinates of the centre of the Gaussian 
relative to the core. Position angle is measured from north through east.}
\tablenotetext{c}{$a$ and $b$ are the FWHM of the major and minor axes of the Gaussian.}
\tablenotetext{d}{Position angle of the major axis measured from north through east.}

\end{center}
\end{table}

\clearpage
\begin{table}
\begin{center}
\caption{Parameters of the 2.3/8.6 GHz VLBI observations from RRFID and VLBA archives}
\begin{tabular}{lllll}
\tableline\tableline
Epoch & Code & Antenna & 2.3GHz beam & 8.6GHz beam \\
 &  & & (mas/mas/deg) & (mas/mas/deg)  \\
\tableline
1995 Oct 13 & RRFID & VLBA  & 13.5/3.97/-15.3 & 2.29/0.955/-1.13  \\
1997 Jan 30 & RRFID & VLBA+6GA  & 8.02/2.88/-4.76 & 2.03/0.745/-5.85\\
1997 Mar 31 & RRFID & VLBA+6GA & 6.55/2.85/-0.75 & 1.71/0.783/1.67\\
1997 May 19 & RRFID & VLBA+6GA  & 8.89/3.20/-6.20 & 2.31/0.849/-5.68\\
1997 Jul 24 & RRFID & VLBA+6GA & 8.41/3.26/-3.10 & 2.24/0.837/-3.52 \\
1997 Sep 8 & RRFID & VLBA+6GA  & 11.0/3.25/-11.5 & 2.29/0.867/-11.6 \\
1997 Dec 17 & RRFID & VLBA+7GA & 9.59/2.83/-13.7 & 2.18/0.718/-10.7 \\
1998 Feb 9 & RRFID & VLBA+7GA & 10.2/2.98/-13.5 & 2.47/0.742/-11.4 \\
1998 Apr 15 & RRFID & VLBA+7GA & 7.45/3.01/-2.55 & 2.04/0.799/-4.24 \\
1998 Jun 25 & RRFID & VLBA+7GA & 10.2/3.28/-3.67 & 2.59/0.712/-3.90 \\
1998 Aug 10 & RRFID & VLBA+7GA &  8.98/2.49/-8.97 & 1.73/0.653/-4.29 \\
1998 Dec 21 & RRFID & VLBA+7GA & 8.60/3.57/1.04 & 2.35/0.777/-0.392 \\
2002 Jan 16 & RRFID & VLBA+8GA & 9.20/2.46/-5.09 & 2.40/0.598/-6.26 \\

2002 Jul 25 & RDV 34 & VLBA+8GA & 6.80/4.45/26.4 & 2.44/1.50/-11.0 \\
2002 Sep 25 & RDV 35 & VLBA+8GA & 6.82/1.91/1.73 & 1.85/0.500/1.45 \\
2003 May 7 & RDV 38 & VLBA+9GA & 2.64/1.83/32.0 & 0.841/0.554/11.3 \\
2003 Jun 19 & RDV 39 & VLBA+8GA & 7.89/4.68/26.5 &  2.01/1.27/20.8 \\
2003 Sep 18 & RDV 41 & VLBA+8GA & 5.61/2.24/-0.761 &  1.67/0.616/-1.48 \\
2003 Dec 17 & RDV 42 & VLBA+9GA & 5.51/2.41/6.17 &  0.808/0.530/51.9 \\
2004 Mar 3 & RDV 43 & VLBA+8GA & 5.96/2.09/-1.33 &  1.69/0.580/-1.38 \\
2004 May 5 & RDV 44 & VLBA+8GA & 6.96/2.80/-11.8 &  1.92/0.787/-12.3 \\

2004 Jul 15 & RDV 45 & VLBA+8GA & 5.50/2.08/-4.83 & 1.54/0.564/-7.10 \\

2005 May 12 & BP 118 & VLBA  & 6.33/2.36/5.05 & 1.75/0.636/2.99 \\
2005 Jun 2 & BP 118 & VLBA  & 6.45/2.49/-0.392 &  1.82/0.658/-2.16 \\
2005 Jul 1 & BP 118 & VLBA  & 6.24/2.55/-0.351 & 1.65/0.720/0.581 \\
\end{tabular}
\tablecomments{VLBA consists of 10 antennas of 25 m. GA
stands for geodetic antennas as described in Tateyama \& Kingham (2004); 
6GA = Gc, Gn, Kk, Mc, On, and Wf; 7GA=6GA+Ny; 8GA=7GA+Ts-Wz-Gn; 
9GA= 8GA+Tc. Gc is Gilcreek (Alaska, 26m), Gn is NRAO20 (Greenbank, 20m),
Kk is Kokee (Hawaii, 20m), Mc is Medicina (Italy, 32 m), 
Ny is Ny Alesund (Norway, 20 m), On is Onsala (Sweden, 20 m), 
Tc is Tigo (Chile, 6m),
Ts is Tsukuba (Japan, 32 m), Wf is Westford (Massachusetts, 18 m), 
and Wz is Wettzell (Germany, 20 m). }

\end{center}
\end{table}

\begin{deluxetable}{lllrrrrrr} 
\tablecolumns{9} 
\tablewidth{0pc} 
\tablecaption{Gaussian models for images of the VLBI observations from RRFID and VLBA archives} 
\tablehead{ 
\colhead{Epoch (MJD)} & \colhead{GHz}  & \colhead{Comp.ID}   & \colhead{$S$(Jy)}    & \colhead{$r$(mas)} & 
\colhead{PA(deg.)}    & \colhead{$a$(mas)}   & \colhead{$b/a$}    & \colhead{$\psi$(deg.)}}
\startdata 

1995 Oct 13 
(50003) & 2.3 & Core & 1.861 & - & - & 0.76 & 0.00 &       76.2\\
   & & C2 & 0.653  &   2.00 & -61.4 & 1.81  & 0.92 & 5.9\\
   & & C1 & 0.290  &   5.47 & -73.3 & 1.52 & 0.83 & 47.4\\
   & 8.6 & Core &   1.544 &  -   & -  & 0.55 & 0.42 & -20.6\\
        & & C3 &    0.697 &  1.07 &   -55.4 & 1.72 &  0.33 &  85.2\\
        & & C1 &    0.157 &  5.27 &   -72.6 & 3.54 &  0.81 &  86.3\\
1997 Jan 30 
(50479) & 2.3 & Core & 1.453 & - & - &  1.26 & 0.72 & -62.2 \\
   & & C2 & 0.395 &     2.25 &  -74.6 & 1.92 &  0.00 &      11.1 \\
   & & C1 & 0.275  &   5.19 &  -75.1 & 2.54 &  0.71 & 24.1 \\
 
  & 8.6 & Core &   1.223 & -       & -     & 0.54 & 0.18 & 19.6\\
        & & C3 &    0.199 & 0.85 &  -47.6 & 0.48 &  0.73 &  48.4\\
        & & C2 &    0.232 & 2.11 &  -70.5 & 1.89 &  0.50 &  -73.4\\
        & & C1 &    0.086 & 5.79 &  -73.9 & 2.43 &  0.88 &  11.0\\
1997 Mar 31 
(50539) & 2.3 & Core & 1.331 & - & - &   1.27 &  0.66 &  -52.1 \\
   & & C2 & 0.517     & 2.28  & -74.3 & 1.93  & 0.78 & -10.4 \\
   & & C1 & 0.227     & 5.51  & -75.3 & 2.17  & 0.57 & 25.0 \\
   & 8.6 & Core &   1.233 & -  & -     & 0.55 &  0.33 &  8.3 \\
        & & C3 &    0.215 & 0.89 &  -46.3 & 0.54 &  0.53 &  -65.5 \\
        & & C2 &    0.208 & 2.32 &  -70.2 & 1.68 & 0.48 &  -66.2 \\
        & & C1 &    0.078 & 6.00 & -73.9 & 2.37 &   0.69 &  27.3 \\
1997 May 19 
 (50588) & 2.3 & Core & 1.587 & - & - &  1.46 &  0.60 & -52.3 \\
   & & C2 & 0.384 &     2.51 &   -74.9 &  1.00 &   0.00  & -5.5 \\
   & & C1 & 0.265 &      5.25  & -76.9 & 3.16  & 0.42 & 4.6 \\
  & 8.6 & Core &   1.148 & - &     -       & 0.59 &        0.34 &  15.9\\
        & & C3 &   0.217 &  0.85 &  -49.0 & 0.63 &  0.74 &  -64.4\\
        & & C2 &   0.225 &  2.34 &  -70.3 & 2.02 &  0.51 &  -74.8\\
        & & C1 &   0.081 &  5.97 &  -73.8 & 2.10 &  0.80 &  37.5\\
1997 Jul 24
        (50654)& 2.3 & Core & - & - & 0.48  & 1.45  & 0.42 & -60.0 \\
   & & C2 & 0.375 &    2.58 &   -73.4 & 1.63  & 0.00     & -1.5 \\
   & & C1 & 0.252 &    5.34 &  -75.3 & 3.10  & 0.35 & 10.4 \\
   & 8.6 & Core&   1.323 & - &     - &  0.56 &     0.37 &  20.4\\
        & & C3 &   0.190 & 0.90 &  -51.0 & 0.46 &  0.07 &  -77.6\\
        & & C2 &   0.234 & 2.40 &  -71.6 & 1.99 &  0.62 &  -79.9\\
        & & C1 &   0.081 & 5.93 &  -74.7 & 2.19 &  0.74 &  23.9\\
1997 Sep 8
        (50700) & 2.3 & Core & 1.56 & - & - & 1.84  & 0.69 & -28.8 \\
   & & C2 & 0.408     & 2.74  & -73.9 & 2.15  & 0.06       & 13.7 \\
   & & C1 & 0.221     & 5.59  & -74.1 & 3.28  & 0.00     & 14.3 \\
  & 8.6 & Core &   1.426 & - &     - &     0.62 &  0.36 &  19.6\\
        & & C3 &   0.201 & 0.91 &  -50.2 & 0.49 &  0.00 &  84.9\\
        & & C2 &   0.252 & 2.44 &  -70.8 & 2.23 &  0.61 &  -74.7\\
        & & C1 &   0.081 & 6.09 &  -73.8 & 2.47 &  0.53 &  16.7\\
1997 Dec 17
        (50800) & 2.3 & Core & 1.538 & - & -  & 1.56  & 0.46 & -58.3\\
   & & C2 & 0.404     & 2.65  & -75.8 & 2.14  & 0.13 & 1.0 \\
   & &  C1 & 0.251     & 5.38  & -75.6 & 1.94  & 0.75 & 17.0 \\
 & 8.6 & Core &   1.390 & - &     - &     0.65 &  0.00  & 22.6\\
        & & C3 &   0.375 & 0.60 &  -53.2 & 1.05 &  0.41 &  -59.7\\
        & & C2 &   0.209 & 2.63 &  -73.7 & 2.03 &  0.63 &  -79.2\\
        & & C1 &   0.085 & 6.07 &  -76.6 & 2.17 &  0.72 &  53.7\\
1998 Feb 9
        (50854) & 2.3 & Core & 1.358 & - & - &  1.46  & 0.65 & -49.2 \\
   & & C2 & 0.451     & 2.75  & -74.9 & 2.08  & 0.62 & -0.2 \\
   & & C1 & 0.189     & 5.70  & -75.6 & 2.16  & 0.00  & 18.7 \\
 & 8.6 & Core &   1.559 & - &     - &     0.63 &  0.08 &  22.7\\
        & & C3 &   0.282 & 0.73 &  -60.6 & 0.90 &  0.43 &  -50.8\\
        & & C2 &   0.200 & 2.75 &  -72.7 & 2.14 &  0.69 &  -80.6\\
        & & C1 &   0.071 & 6.22 &  -76.5 & 1.59 &  0.94 &  12.4\\

1998 Apr 15 
        (50919) & 2.3 & Core &  1.386 & - & -  & 1.19 &  0.66 & -67.9 \\
   & & C2 & 0.428     & 2.57  & -74.4 & 1.86  & 0.46 & 22.1 \\
   & & C1 & 0.258     & 5.39  & -74.4 & 2.18  & 0.63 & 53.8 \\
  & 8.6 & Core &   1.731 & - &     - &     0.62 &  0.13 &  20.6\\
        & & C3 &   0.272 & 0.79 &  -64.9 & 0.88 &  0.53 &  -30.4\\
        & & C2 &   0.208 & 2.77 &  -74.3 & 2.26 &  0.63 &  -80.6\\
        & & C1 &   0.070 & 6.24 &  -75.2 & 1.90 &  0.58 &  22.5\\

1998 Jun 25
        (50989) & 2.3 & Core & 1.526  & - & - & 1.84 & 0.79 & -47.7 \\
   & & C2 & 0.347     & 3.30  & -68.3 & 2.74  & 0.38 & 1.5 \\
   & & C1 & 0.176     & 5.73  & -79.4 & 4.93  & 0.00  & -3.8 \\
  & 8.6 & Core &   1.714 & - &     - &     0.61 &  0.00 &  21.4\\
        & & C3 &   0.427 & 0.59 & -72.0 & 1.33 &  0.25 &  -59.8\\
        & & C2 &   0.167 & 2.86 &  -75.2 & 2.11 &  0.56 &  -79.6\\
        & & C1 &   0.057 & 6.21 &  -76.7 & 1.95 &  0.47 &  32.2\\
        & & C0? &    0.016 & 9.67 &  -76.4 & 1.00 &  0.00 &  26.4\\
1998 Aug 10
        (51036) & 2.3 & Core & 1.463 & - & -  & 1.27 &   0.41 &  -79.7 \\
   & & C2 & 0.359    & 2.69 & -74.2 & 1.55  & 0.00  & 28.2 \\
   & & C1 & 0.250    & 5.34 & -75.4 & 2.11  & 0.51 & 56.7 \\
  & 8.6 & Core &   1.863 & - &     - &     0.59 &  0.20 &  19.5\\
        & & C3 &   0.296 & 0.86 &  -69.4 & 0.93 &  0.53 &  -25.8\\
        & & C2 &   0.208 & 2.84 &  -75.3 & 2.40 &  0.53 &  -88.3\\
        & & C1 &   0.074 & 6.30 &  -77.0 & 1.47 &  0.95 &  64.1\\
1998 Dec 21
        (51169) & 2.3 & Core & 1.437 & - & - & 1.69 & 0.50 & 86.2 \\
   & & C2 & 0.344    & 2.95 &  -71.6 & 1.51 & 0.54 & 52.0 \\
   & & C1 & 0.202    & 5.70 &  -76.5 & 2.83 &  0.29 & 29.0 \\
  & 8.6 & Core&  2.160 & - &     - &     0.56 &  0.51 &  15.8\\
        & & C3 &   0.428 & 0.86 &  -68.5 & 1.68 &  0.43 &  -78.9\\
        & & C2 &   0.086 & 3.42 &  -80.3 & 1.60 &  0.00 &  17.8\\
        & & C1 &   0.088 & 6.30 &  -77.7 & 4.17 &  0.04 &  24.0\\
2002 Jan 16
        (52291)& 2.3 & Core & 1.211 & - & -  & 1.72  & 0.86 & -43.7 \\
   & & C2 & 0.193     & 3.19  & -66.0 & 1.70  & 0.00  & 4.4 \\
   & & C1 & 0.503     & 5.87  & -80.4 & 2.09  & 0.65 & -3.5 \\
 & 8.6 & Core &   1.202 & - &     - &     0.80 &  0.32 &  5.5\\
        & & C4 &    0.197 & 1.48 &  -59.8 & 1.33 &  0.73 &  5.8\\
        & & C3? &   0.015 & 3.32 &  -70.5 & 3.15 &  0.00  &  4.1\\
        & & C2 &   0.036 & 4.26 &  -83.4 & 2.94 &  0.00 &   76.4\\
        & & C1 &   0.203 & 6.53 &  -77.0 & 2.06 &  0.42 &  1.9\\
2002 Jul 25
        (52481)& 2.3 & Core & 1.000  & - & - & 1.82  & 0.56 & -76.4 \\
 & & C1 &    0.463 &      5.60 &   -79.7 & 2.38 &  0.69 & 82.2 \\
 & 8.6 & Core&   1.979 & - &     - &     0.86 &  0.56 &  13.6\\
        & & C4 &   0.133 & 2.35 &  -66.9 & 1.35 &  0.47 &  16.3\\
        & & C2&    0.072 & 5.04 &  -70.0 & 1.79 &  0.00 &  21.8\\
        & & C1 &   0.220 & 6.61 &  -78.3 & 1.58 &  0.36 &  12.1\\
2002 Sep 25
        (52543)& 2.3 & Core & 1.201  & - & & 1.45  & 0.23 & -15.2 \\
   & & C4 & 0.390     & 2.43  & -79.2 & 3.73  & 0.25 & -50.8 \\
   & & C1 & 0.507     & 6.14  & -80.9 & 1.79  & 0.00     & -41.1 \\
 & 8.6 & Core&   1.794 & - &     - &     0.81 &  0.30 &  -0.1\\
        & & C3 &   0.286 & 1.29 &  -80.9 & 2.51 &  0.56 &  -6.2\\
        & & C2 &   0.079 & 5.01 &  -78.3 & 3.54 &  0.38 &  17.8\\
        & & C1 &   0.218 & 6.69 &  -79.5 & 1.49 &  0.53 &  13.0\\
2003 May 7
        (52767) & 2.3 & Core & 1.384 & - & - &  1.97 &  0.43 & 1.5 \\
   & & C4 & 0.566     & 1.45  & -74.6 & 4.63  & 0.57 & -27.2 \\
   & & C1 & 0.766     & 6.14  & -81.2 & 2.48  & 0.56  & 7.6 \\
  & 8.6 & Core &   2.088 & - &     - &     0.83 &  0.36 &  -0.1\\
        & & C4 &   0.176 & 2.40 &  -69.5 & 4.14 &  0.00 &  85.9\\
        & & C3 &   0.122 & 1.04 &  -94.7 &  0.46 &  0.00 &  89.2\\
        & & C1 &   0.291 & 6.70 &  -78.8 & 1.41 &  0.74 &  14.3\\
2003 Jun 19
        (52809)& 2.3 & Core &  2.362 & - & - & 3.77  & 0.57 & -72.4 \\
 & & C4 &  0.688 &    6.56 &  -78.7 &  7.95 &  0.00 &      8.1 \\
 & 8.6 & Core&   2.113 & - &     - &     0.87 &  0.47 &  -8.2\\
        & & C4 &   0.181 & 3.07 &  -79.0 & 3.37 &  0.00 &  -70.5\\
        & & C1 &   0.401 & 6.47 &  -81.3 & 1.79 &  0.69 &  -4.6\\

2003 Sep 18
        (52886)& 2.3 & Core &  1.108 & - & - &   2.06  & 0.25 & -20.6 \\
   & & C4 & 0.602     & 1.66  & -80.5 & 5.13  & 0.52 & -43.0 \\
   & & C1 & 0.702     & 6.34  & -81.0 & 2.79  & 0.46 & -17.5 \\
 & 8.6 & Core &   2.027 & - &     - &     0.74 &  0.43 &  -1.8\\
        & & C2 &   0.209 & 6.22 &  -73.9 & 3.19 &  0.00 &  -72.5\\
        & & C3 &   0.408 & 0.797 & -94.8 &  3.26 &  0.00 &  -50.9\\
        & & C1 &   0.186 & 6.78 &  -82.2 & 1.41 &  0.49 &  -30.7\\

2003 Dec 17
        (52991)& 2.3 & Core & 1.199  & - & - & 1.18  & 0.00     & -47.9 \\
   & & C4 & 0.295     & 2.61  & -70.7 & 3.01  & 0.64 & -34.7 \\
   & & C1 & 0.612     & 6.38  & -80.8 & 2.04  & 0.53 & -17.0 \\
 & 8.6 & Core & 2.229 & - &       - &     0.62 &  0.53  &  8.7\\
        & & C4 &   0.060 & 3.03 & -74.2 & 2.39 & 0.27 & 40.3\\ 
        & & C3 &   0.185 & 0.76 & -61.9 & 1.81 & 0.00 & 40.7\\
        & & C2 &   0.058 & 6.63 & -70.2 & 0.59 & 0.00 & 89.5\\ 
        & & C1 &   0.249 & 6.77 & -82.0 & 1.42 &  0.54 & -26.3\\
2004 Mar 3
        (53067)& 2.3 & Core & 1.068 & - & - &   1.37 &   0.00 &  -27.1 \\
   & & C4 & 0.286      & 2.24  & -69.9 & 3.92  & 0.12 &  -60.0 \\
   & & C1 & 0.559     & 6.41  & -81.0 & 1.64  & 0.55 & -30.0 \\
 & 8.6 & Core & 1.768 & - &       - &     0.49 &  0.46  &  26.8\\
        & & C3 &   0.147 & 0.71 & -40.2 & 0.75 & 0.00 & 46.5\\
        & & C1 &   0.251 & 6.87 & -79.4 & 1.59 &  0.59 & 6.6\\
2004 May 5
        (53131)& 2.3 & Core &  1.829 & - & - & 1.86  & 0.55 & -25.4 \\
   & & C4 & 0.128     & 3.10 & -71.3 & 13.30  & 0.00     &  -18.6 \\
   & & C1 & 0.883     & 6.07 & -82.9 & 2.70 &  0.66 & -12.9 \\
 & 8.6 & Core & 2.364 & - &       - &     0.51 &  0.00     &  23.1\\
        & & C4 &   0.171 & 4.53 & -78.8 & 6.80 & 0.42 & -37.5\\        
        & & C3 &   0.109 & 1.56 & -84.8 & 2.01 & 0.22 & -19.0\\
        & & C1 &   0.349 & 6.80 & -82.4 & 1.45 & 0.59 & 13.2\\ 
2004 Jul 15
        (53201)& 2.3 & Core & 1.351 & - & - &  1.14 & 0.70 & -9.7 \\
   & & C4 & 0.352     & 2.29  & -73.7 & 4.00  & 0.52 & -49.0 \\
   & & C1 & 0.694     & 6.48  & -81.2 & 2.08  & 0.61 & 2.9 \\
 & 8.6 & Core &   2.481 & - &     - &     0.52 &  0.44 &  21.0\\
        & & C4 &   0.128 & 4.74 &  -78.7 & 3.68 &  0.92 &  -6.7\\
        & & C5? &   0.356 & 0.59 &  -56.4 & 0.61 &  0.00 &  53.4\\
        & & C3 &   0.188 & 1.16 &  -97.2 &  3.15 &  0.11 &  -39.0\\
        & & C1 &   0.332 & 6.89 &  -80.7 & 1.43 &  0.70 &  1.4\\

2005 May 12
        (53493)& 2.3 & Core & 1.473 & - & -  & 1.28  & 0.81 & -61.9 \\
   & & C4 & 0.121     & 3.01  & -70.1 & 0.42 & 1.00  & -14.4 \\
   & & C1 & 0.641     & 6.29  & -81.0 & 2.58  & 0.46 & 3.7 \\
 & 8.6 & Core &  2.098 & - &       - &    0.76 &  0.32  &  9.65\\
        & & C4 & 0.201  & 5.69 & -85.0 & 6.48 & 0.49 & 63.5\\
        & & C3 & 0.467 & 0.91 & -103.3 & 2.15 & 0.42 & 27.3\\
        & & C1 & 0.167 & 7.02 & -81.4 & 1.58 & 0.00 & -51.0\\

2005 Jun 2
        (53523)& 2.3 & Core & 1.602 & - & - &  1.23  & 0.85 & 9.5 \\
   & & C4 & 0.354 &      2.42 &  -71.7 & 3.06  & 1.00 &   -90.0 \\
   & & C1 & 0.748     & 6.44  & -80.9 & 2.37  & 0.49 & 9.0 \\
 & 8.6 & Core &  2.268  & - &       - &    0.81 &  0.38  &  2.9\\
        & & C4 & 0.069 & 5.87 & -92.1 & 4.05 & 0.00 & 34.6\\
        & & C3 & 0.269 & 0.69 & -65.1 & 0.56 & 0.34 & 87.6\\
        & & C1 & 0.277 & 6.88 & -80.7 & 1.62 & 0.59 & -2.5\\

2005 Jul 1
        (53554)& 2.3 & Core & 1.752 &  - & - &  1.27 &  0.84 & -87.8 \\
   & & C4 & 0.216     & 3.01  & -68.4 & 1.48  & 1.00  & -99.6 \\
   & & C1 & 0.765     & 6.34  & -81.1 & 2.08  & 0.64 & 16.0 \\
 & 8.6 & Core &  2.262  & - &       - &    0.79 &  0.42  &  4.1\\
        & & C3 & 0.181 & 0.73 & -83.2 & 1.90 & 0.00 & 3.2\\
        & & C1 & 0.367 & 6.78 & -81.0 & 1.98 & 0.57 & 6.3\\

\enddata 
\end{deluxetable}

\clearpage
\begin{table}
\begin{center}
\caption{Parameters of the 2.3/8.6 GHz band VLBI maps shown in Fig. 3-5.}
\begin{tabular}{lclcl}
\tableline\tableline
 & 2.3GHz peak & 2.3GHz contour &  8.6GHz peak & 8.6GHz contour \\
Epoch &  (Jy/beam) & \% of peak & (Jy/beam) & \% of peak \\
\tableline
2002 Jul 25   & 1.07 & -2 2 4 8 16 32 64 & 1.77 & -0.4 0.4 0.8 1.6 3.2 6.4 12.8 25.6   \\
2002 Sep 25   & 1.18 & -1 1 2 4 8 16 32 64  & 1.53 & -1 1 2 4 8 16 32 64  \\
2003 May 7  & 1.09 & -1.5 1.5 3 6 12 24 48 96 & 1.26 & -1 1 2 4 8 16 32 64   \\
2003 Jun 19 & 1.8 & -1 1 2 4 8 16 32 64 & 1.93 & -1.5 1.5 3 6 12 24 48 96 \\
2003 Sep 18  & 1.16 & -1.5 1.5 3 6 12 24 48 96 & 1.64 & -0.75 0.75 1.5 3 6 12 24 48 96  \\
2003 Dec 17   & 1.22 & -1 1 2 4 8 16 32 64 & 1.03 & -1.5 1.5 3 6 12 24 48 96   \\
2004 Mar 3   & 1.09 & -0.75 0.75 1.5 3 6 12 24 48 96 & 1.55 & -0.5 0.5 1 2 4 8 16 32 64   \\
2004 May 5  & 1.66 & -0.5 0.5 1 2 4 8 16 32 64 & 2.51 & -0.5 0.5 1 2 4 8 16 32 64 \\
2004 Jul 15 & 1.29 & -0.75 0.75 1.5 3 6 1 12 24 48 96 & 1.87 & -0.5 0.5 1 2 4 8 16 32 64   \\
2005 May 12   & 1.56 & -1 1 2 4 8 16 32 64 & 1.83 &  -0.75 0.75 1.5 3 6 12 24 48 96  \\
2005 Jun 2   & 1.56 & -1 1 2 4 8 16 32 64 & 1.96 & -0.5 0.5 1 2 4 8 16 32 64  \\
2005 Jul 1   & 1.59 & -0.5 0.5 1 2 4 8 16 32 64 & 1.90 & -0.5 0.5 1 2 4 8 16 32 64 \\
\tableline
\end{tabular}
\end{center}
\end{table}

\clearpage
\begin{table}
\begin{center}
\caption{Proper motions modeled from multi epoch VLBI data at 8.6 GHz}
\begin{tabular}{lcccl}
\tableline\tableline
Component & $t_0$(Modified J.D.)\tablenotemark{a} & $t_0$(Epoch)\tablenotemark{a}  & $\mu$(mas/yr) & $\beta_{app}$ \\
\tableline
C1  &    32470 $\pm$ 1866 & 1947.79 $\pm$ 5.11 & 0.12 $\pm$ 0.01 & 2.0 $\pm$ 0.2   \\
C2  &    49185 $\pm$ 75.3 & 1993.54 $\pm$ 0.21 & 0.58 $\pm$ 0.02 & 9.3 $\pm$ 0.3  \\
C3  &    24942 $\pm$ 26268 & 1927.16 $\pm$ 71.97 & 0.01 $\pm$ 0.01 & 0.2 $\pm$ 0.2   \\
C4  &    51977 $\pm$ 42.9 & 2001.18 $\pm$ 0.12 & 1.35 $\pm$ 0.05 & 21.6 $\pm$ 0.8  \\
\tableline
\end{tabular}
\tablenotetext{a}{Zero Separation Time}
\end{center}
\end{table}







\begin{thebibliography}{}

\bibitem[Agudo et al. (2001)]{agudo2001} 
   Agudo, I. et al. 2001, \apj, 549, L183

\bibitem[Aller (1970)]{aller70} 
   Aller, H.D. 1970, \apj, 161,19

\bibitem[Aller, H.D. et al.(1985)]{aller85} 
   Aller, H.D., Aller, M.F., Latimer, G.E., \& Hadge, P.E. 1985, \apjs,
    59, 513

\bibitem[Alberdi et al. (2000)]{alberdi00}
    Alberdi, A., G\'{o}mez, J.L., Marcaide, J.M., Marscher, A.P.,
\& P\'{e}rez-Torres, M.A. 2000, \aap, 361, 529

\bibitem[Bach et al. (2005)]{bach05} 
   Bach, U. et al. 2005 \aap,    433, 815


\bibitem[Blandford \& Rees(1974)]{blandford74} 
   Blandford, R. D., \& Rees, M. J. 1974, \mnras,    169, 395

\bibitem[Britzen (2005)]{britzen05} 
   Britzen, S. et al. 2005 \mnras, 362,966


\bibitem[Conway \& Murphy (1993)]{conway93} 
   Conway, J.E. \& Murphy D.W. 1993, \apj, 411, 89

\bibitem[D'Addario,L.R (1989)]{d'addario89} 
  D'Addario, L.R. 1989, in ASP cond. Ser. 6, Synthesis Imaging 
  in Radio Astronomy, Perley, R.A., Schwab, F.R., \& Bridle, A.H., 59

\bibitem[Fey \& Charlot (1997)]{fey97}
    Fey, A. L., \& Charlot. P. 1997, \apjs, 111, 95

\bibitem[Flomalont et al. (2000)]{fomalont00}
    Fomalont, E.B., Frey, S., Paragi, Z., Gurvits, L.I., Scott, W.K., 
Taylor, A.R., Edwards, P.G., \& Hirabayashi, H. 2000, \apjs, 111, 95


\bibitem[Gabuzda \& G\'{o}mez (2001)]{gabuzda01}
    Gabuzda, D. C., G\'{o}mez, J.L. 2001, \mnras,    320, L49

\bibitem[Gabuzda, Pushkarev \& Cawthorne(2000)]{gabuzda00}
    Gabuzda, D. C., Pushkarev, A. B., \& Cawthorne, T. V. 2000, \mnras,
    319, 1109

\bibitem[G\'{o}mez et al. 2001]{gomez01}
    G\'{o}mez, J.L., et al. 2001, \apj, 561, L161

\bibitem[Gordon (2000)]{gordon00}
    Gordon, D. 2000, in International VLBI Service for Geodesy and Astrometry:
    2000 General Meeting Proceedings, 361

\bibitem[Himwich (1988)]{himwich88}
    Himwich, W.E. 1988, in The impact of VLBI on astrophysics and geophysics; 
  Proceedings of the 129th IAU Symposium,   
  Dordrecht, Kluwer Academic Publishers, 357

\bibitem[Horiuchi et al. (2004)]{horiuchi04}
    Horiuchi, S., et al. 2004, \apj, 616, 110

\bibitem[Jorstad et al. (2004)]{jorstad04}
    Jorstad, S.G. et al. 2004, \apj, 127, 3115

\bibitem[Jenet et al. (2004)]{jene04}
    Jenet, F.A., Lommen, A., Lason, S.L., \& Wen, L. 2004, \apj, 606, 799


\bibitem[Kellermann(1998)]{kellermann98}
    	Kellermann, K.I., Vermeulen, R. C., Zensus, J. A., \& Cohen, M. H.
         1998, \aj, 115, 1295

\bibitem[Kellermann(2004)]{kellermann04}
    Kellermann, K.I. et al. 2004, \apj, 609, 539

\bibitem[Kollgaart(1994)]{kollgaart94}
    Kollgaard, R.I. 1994, Vistas in Astronomy, 38, 29

\bibitem[Kovalev et al.(2005)]{kovalev05}
    Kovalev et al. 2005, \aj, 130, 2473

\bibitem[Koyama et al. (2003)]{koyama03}
  Koyama, Y., Kondo, T., Nakajima, J., Sekido, M., \& Kimura, M. 2003,
 in ASP Conf. Series 306, New Technologies in VLBI, 
 ed. by Minh, Y.C. (San Francisco: APS), 135

\bibitem[K\"{u}hr \& Schmid (1990)]{kuhr90}
    K\"{u}hr, H. \& Schmidt, G.D. 1990, \aj, 99, 1


\bibitem[Lister \& Homan (2005)]{lister05}
    Lister, M.L. \& Homan, D.C. 2005, \aj, 130, 1389


\bibitem[Lovell (2000)]{lovell00}
    Lovell, J. 2000, Astrophysical Phenomena Revealed by Space VLBI, ed. 
    Hirabayashi, H., Edwards, P.G. \& Murphy, D.W. 
    (Inst. of Space and Astronautical Sci.), 301,
    http://www.atnf.csiro.au/people/Jim.Lovell/difwrap/

\bibitem[Migliari, Fender, \& M\'{e}nder (2002)]{migliary02}
    Migliari, S., Fender, R., M\'{e}ndez, M. 2002, Science, 297, 1673


\bibitem[Perley (1982)]{perley82} 
    Perley, R.A. 1982, \aj, 87, 859



\bibitem[Roberts (1997)]{roberts97} 
    Roberts, P.P. 1997, \aap, Supplement Series, 126, 379

\bibitem[Ritakari \& Mujunen (2002)]{ritakari02} 
    Ritakari, J. \& Mujunen, A. 2002, in International VLBI Service 
    for Geodesy and Astrometry: General Meeting Proceedings
     (National Technical Information Service), 128



\bibitem[Scott et al.(2004)]{scott04} 
    Scott, W.K. et al. 2004, \apjs, 155, 33

\bibitem[Shepherd (1997)]{shepherd97} 
    Shepherd, M.C. 1997, in ASP Conf. Series 125, Astronomical Data Analysis 
    Software System VI, ed. Hunt, G. and Payne, H.E. (San Francisco: ASP), 77

\bibitem[Stickel et al.(1989)]{stickel89} 
    Stickel, M., Fried, J. W., \& K\"{u}hr, H. 1989, \aap Supplement Series, 80, 103



\bibitem[Takeuchi (2004)]{takeuchi04}
    Takeuchi, H. 2004, in IVS NICT Technology Development Center News
    (National Institute of Information and Communications Technology, Tokyo), 24, 9

\bibitem[Tateyama et al.(2002)]{tateyama02} 
    Tateyama, C.E., Kingham, K.A., Kaufman, P., de Lucena, A.P. 2002, \apj, 573, 496

\bibitem[Tateyama \& Kingham (2004)]{tateyama04} 
    Tateyama, C.E. \& Kingham, K.A. 2004, \apj, 608, 149

\bibitem[Tngay et al.(2001)]{tingay01}
    Tingay, S. et al. 2001, \aj, 122, 1697

\bibitem[Tngay et al.(2003)]{tingay03}
    Tingay, S. et al. 2003, \pasj, 55, 351

\bibitem[Ulvestat et al.(1981)]{ulvestat81}
    Ulvestad, J., Johnston, K., Perley R., \& Fomalont, E. 1981, \aj, 86, 1010

\bibitem[Urry \& Padovani (1995)]{urry95}
    Urry, L.M., \& Padovani, P. 1995, PASP, 107, 803 


\bibitem[West (2004)]{valtaojy00}
    West, C.J. 2004, Masters thesis, Swinburne University of Technology

\bibitem[Wilson et al.(1992)]{wilson92}
    Wilson, W.E., Davis, E.R., Loone, D.G. \& Brown, D.R. 1992, 
    Jounal of Electrical and Electronics Engineering, Australia, 12, 187


\end{thebibliography}
\end{document}